\documentclass[amsmath,amssymb,aps,prx,twocolumn,superscriptaddress]{revtex4-1}

\usepackage{graphicx}% Include figure files
\usepackage{dcolumn}% Align table columns on decimal point
\usepackage{bm}% bold math
\usepackage[T1]{fontenc}
\usepackage{lmodern}
\usepackage{graphicx}
\usepackage{floatrow}%
\usepackage{cancel}

\begin{document}

\title{Performance of a quantum heat engine at strong reservoir coupling}

\author{David Newman}
\affiliation{Department of Physics, Imperial College London, London, SW7 2AZ, UK}

\author{Florian Mintert}
\affiliation{Department of Physics, Imperial College London, London, SW7 2AZ, UK}

\author{Ahsan Nazir}
\affiliation{Photon Science Institute and School of Physics and Astronomy, The University of Manchester, Oxford Road, Manchester, M13 9PL, UK}

\date{\today}

\begin{abstract}
We study a quantum heat engine at strong coupling between the system 
and the thermal reservoirs. Exploiting a collective coordinate mapping, we incorporate system-reservoir correlations into a consistent 
thermodynamic analysis, thus circumventing the usual restriction to weak coupling and vanishing correlations.  
We apply our formalism to the example of a quantum Otto cycle, demonstrating that the performance of the engine is diminished in the strong coupling regime with respect to its weakly coupled counterpart, producing a reduced net work output and operating at a lower energy conversion efficiency. We identify costs imposed by sudden decoupling of the system and reservoirs around the cycle  
as being primarily responsible for 
the diminished performance, and define an alternative operational procedure 
which can partially recover 
the work output and 
efficiency. More generally, the collective coordinate mapping holds considerable promise for wider studies of thermodynamic systems beyond weak reservoir coupling.
\end{abstract}


\maketitle

\section{Introduction}

Heat engines played a major role in formulating  
the laws of thermodynamics. For example, the second law can be understood in terms of the efficiency of an ideal Carnot cycle \cite{Carnot:1824aa,Blundell:2010aa}. Recently, significant effort has been invested into studying 
quantum mechanical analogues of these engines as 
%one
an approach to establishing 
whether the same laws 
apply also to quantum systems; see Refs.~\cite{doi:10.1146/annurev-physchem-040513-103724,Gelbwaser-Klimovsky2015,arXiv:1508.06099,e15062100} for some contemporary reviews. Though quantum engines that have been designed  
to operate in standard 
cycles seem to respect the laws of thermodynamics~\cite{Skrzypczyk:2014aa,PhysRevLett.2.262,PhysRevE.76.031105,PhysRevLett.93.140403}, it has been shown 
that in certain situations quantum effects can be used to enhance their performance \cite{Scully862,PhysRevLett.88.050602}, and in some cases even violate classical thermodynamic bounds \cite{PhysRevLett.112.030602,0295-5075-88-5-50003}. However, these treatments usually 
involve circumstances beyond the scope for which the laws apply, for example non-thermal baths \cite{PhysRevLett.112.030602,0295-5075-88-5-50003} or systems out of equilibrium \cite{EPL2014Lutz}. 

Typically, quantum mechanical models of heat engines are restricted to the weak coupling regime \cite{PhysRevE.87.012140,Gelbwaser-Klimovsky2015, doi:10.1146/annurev-physchem-040513-103724, PhysRevE.76.031105, 1367-2630-8-5-083, PhysRevLett.112.030602, Friedenberger2015}. That is, they assume 
the interaction strength between the working system and each reservoir to be negligible in comparison to 
their respective self-energies. During  
processes in which the system is coupled to a reservoir  
the total state 
may then be approximated as remaining
%separable,
a product state, with no correlations generated between the two.  As well as significantly simplifying the analysis, this is advantageous as it makes  
distinguishing energy flows in terms of heat and work less problematic \cite{PhysRevLett.116.020601}. However, it is of both fundamental and practical importance 
to understand whether and how thermodynamic treatments  
can be modified to apply beyond such simplifying assumptions. 
For example, the strong coupling regime is experimentally accessible in nanoscale devices \cite{PhysRevB.88.195414,PhysRevB.87.115419,PhysRevB.85.140506,PhysRevLett.110.017002,PhysRevLett.111.243602,0953-8984-28-10-103002,PhysRevLett.113.097401} and exciting technological implications of quantum heat machines have been proposed, such as in laser cooling~\cite{Vogl:2009aa,PhysRevA.91.023431}. This motivates the requirement for a greater understanding of heat engines which operate 
under conditions of strong reservoir coupling~\footnote{We consider any situation in which the system-reservoir factorisation assumption fails to define the strong coupling regime.}.

Though considerable attention has recently been focused on identifying consistent definitions of heat, work, and internal energy in the strong coupling regime \cite{PhysRevB.90.075421,PhysRevLett.116.020601,1367-2630-17-4-045030,PhysRevLett.114.080602},  
and on formulating the second law by studying strong coupling versions of quantum fluctuation relations \cite{PhysRevLett.102.210401,RevModPhys.83.771,Hanggi:2015aa}, or entropy and entropy production \cite{Horhammer2008,1367-2630-12-1-013013,PhysRevLett.107.140404,1742-5468-2013-04-P04005}, a consensus on a consistent approach to 
analysing thermodynamic cycles beyond weak reservoir coupling is arguably still lacking. 
In fact, there are few studies of heat engines in this regime 
to date, with some exceptions being Refs.~\cite{doi:10.1021/acs.jpclett.5b01404, arXiv:1602.01340,e18050186} that analyse continuously coupled engines, and Ref.~\cite{1367-2630-16-12-125009}, which considers a general work extraction process from a single temperature reservoir, rather than a full heat engine cycle.

In this article, we study a quantum heat engine operating in a discrete stroke thermodynamic (Otto) cycle between two thermal reservoirs, 
close in spirit to its classical counterpart, though without the usual restriction to weak reservoir coupling and vanishing system-reservoir correlations. 
By generalising the energetic 
analysis of the cycle strokes to 
account for both the system and reservoirs, including sudden coupling and decoupling steps, we find that 
strong reservoir coupling 
acts to diminish the engine's performance compared to a standard weak coupling treatment. This is due primarily to 
the work cost incurred 
in turning off such interactions, 
which we show can be mitigated by an adiabatic decoupling procedure 
that partially recovers the engine's output.

To facilitate thermodynamic calculations in the strong coupling regime 
we apply 
a collective coordinate mapping (see below). This was recently shown to predict the dynamics and equilibrium states of a quantum spin strongly coupled to a bosonic environment---the spin-boson model \cite{RevModPhys.59.1,weissbook,nitzanbook,Breuer:2002aa,Schlosshauer:2007aa}---in very close agreement with accurate numerical simulations \cite{PhysRevA.90.032114}, and also independently put forward in Ref.~\cite{arXiv:1602.01340} to analyse a continuously coupled engine. As we shall show, the collective coordinate mapping 
enables us to redefine the boundary between our system and its thermal reservoirs, making calculations of energy flows tractable without recourse to vastly simplifying weak coupling assumptions that neglect the generation of system-reservoir correlations. 
Furthermore, it allows us to retain a description in terms of reduced thermal states even within the strong coupling regime, albeit defined on an enlarged state space, such that our engine cycle calculations proceed in very close analogy to standard thermodynamic approaches.

\section{Otto cycle model \label{OttoModel}}

We shall focus on the example of a heat engine operating in an Otto cycle. Quantum Otto cycles have previously been studied in the weak coupling regime \cite{Gelbwaser-Klimovsky2015, doi:10.1146/annurev-physchem-040513-103724, PhysRevE.76.031105, 1367-2630-8-5-083, PhysRevLett.112.030602, Friedenberger2015} with various results ranging from those analogous to classical thermodynamic bounds \cite{PhysRevE.76.031105}, to interesting violations thereof~\cite{PhysRevLett.2.262}. An advantage of the Otto cycle 
is that it allows energetic changes to be distinguished by means of separate strokes where either work is extracted from (or done on) the system, or energy is exchanged between the system and the reservoirs. 

\begin{figure}
\includegraphics[width=0.99\textwidth]{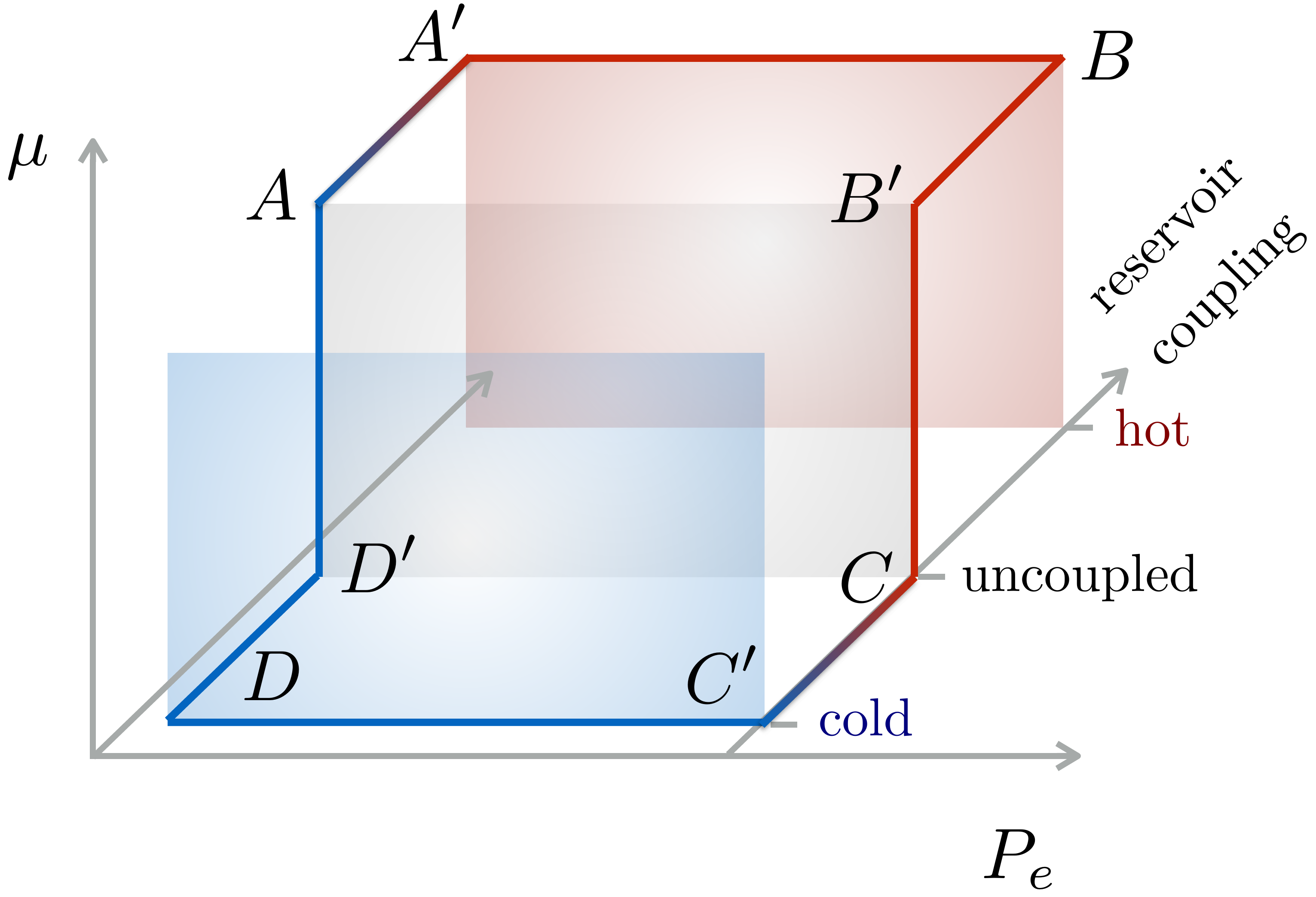}
\caption{Quantum Otto cycle for a TLS with ground state $|g\rangle$ and excited state $|e\rangle$. The vertical axis represents the TLS splitting $\mu$ and the horizontal axis represents the population of the excited state $P_e$. The shaded planes indicate hot and cold reservoirs at temperatures $T_h$ and $T_c$, respectively. It is standardly a four stroke cycle: isochoric thermalisation $A' \rightarrow B$, isentropic expansion $B' \rightarrow C$,  isochoric thermalisation $C' \rightarrow D$, and isentropic compression $D' \rightarrow A$. However, here we expand the cycle to explicitly include energetic contributions from coupling and decoupling the system and reservoirs: coupling to the hot reservoir $A\rightarrow A'$, decoupling from the hot reservoir $B\rightarrow B'$, coupling to the cold reservoir $C\rightarrow C'$, and decoupling from the cold reservoir $D\rightarrow D'$.
\label{quantum-otto-cycle}}
\end{figure}

We consider a quantum system 
with self-Hamiltonian $H_S$, which may be coupled to and decoupled from two heat reservoirs, one at the hot temperature $T_h$ and the other at the cold temperature $T_c$. 
The protocol for our Otto cycle is schematically depicted in Fig.~\ref{quantum-otto-cycle}. It consists of four strokes that also include system-reservoir coupling and decoupling steps:
\begin{itemize}
  \item $A'\rightarrow B$: At point $A'$ the system is coupled to the hot thermal reservoir at temperature $T_h$ and then allowed to reach a steady state (point $B$) while its self-Hamiltonian remains fixed. The classical analogue of this stroke is referred to as the hot isochore. The system is then decoupled from the hot reservoir ($B\rightarrow B'$). 
In standard treatments of the cycle, there is no energy exchange associated with the decoupling step and so it is typically 
ignored.
  \item $B'\rightarrow C$: From point $B'$, the system does not interact with either reservoir. Along the stroke, referred to as isentropic expansion, its self-Hamiltonian is changed from $H_{S}^{B}$ to $H_{S}^{C}$. Once at point $C$, the system is coupled to the cold reservoir, point $C'$, ready for the next stroke.
  \item $C'\rightarrow D$: At point $C'$, the system is allowed to interact with the cold thermal reservoir at temperature $T_c$ and reaches a steady state (point $D$) while its self-Hamiltonian remains fixed. Classically, this stroke is referred to as the cold isochore. The system is then decoupled from the cold reservoir to reach point $D'$.
  \item $D'\rightarrow A$: From point $D'$, the system does not interact with either reservoir. Its self-Hamiltonian is changed from $H_{S}^{D}$ ($=H_{S}^{C}$) to $H_{S}^{A}$ ($=H_{S}^{B}$) during the stroke, 
  known as isentropic compression. The system is then coupled to the hot reservoir, reaching point $A'$, and the cycle proceeds once more.
  \end{itemize}
  
In order to perform the cycle analysis, 
we shall study a model consisting 
of a two-level system (TLS) interacting sequentially with two harmonic oscillator reservoirs, $R_h$ (hot) and $R_c$ (cold). This model, typically referred to as the spin-boson model, is paradigmatic in the study of dissipation in quantum systems \cite{RevModPhys.59.1}, and has been applied to case studies such as semiconductor quantum dots, spins in magnetic fields, superconducting circuits and decoherence in biological systems \cite{Liu285,PhysRevLett.113.036801,PhysRevB.85.140506,PhysRevLett.110.017002,PhysRevLett.111.243602,0953-8984-28-10-103002,PhysRevLett.114.196802,Shevchenko20101,Gilmore2005}. The self-Hamiltonian for the TLS is (we set $\hbar=1$ throughout)
\begin{equation}\label{Hs}
H_{S}(t)  =  \frac{\mu(t)}{2}I+\frac{\epsilon(t)}{2}\sigma_{z}+\frac{\Delta(t)}{2}\sigma_{x},
\end{equation}
where $\epsilon(t)$ represents the TLS bias, $\Delta(t)$ the tunnelling matrix element, $\sigma_{x,z}$ denote the usual Pauli matrices, and $I$ is the identity. The eigenstates of $H_S$ are associated with eigenvalues $0$ and $\mu(t)$, with splitting given by 
\begin{equation}
\mu(t)=\sqrt{\epsilon^{2}(t)+\Delta^{2}(t)}.
\label{mu}
\end{equation}
The first term in Eq.~(\ref{Hs}) thus provides a time-dependent shift of the system energy scale.
Due to the periodicity of the cycle, it does not affect the work output or efficiency so that it can be chosen at will. We shall therefore use it to arrive at an unambiguous definition of positive (negative) work 
as that done on (by) the system during the relevant isentropic strokes. 

We assume that the TLS couples to each reservoir via oscillator position, which may be expressed along with the reservoir self-Hamiltonian in the general form 
\begin{equation}\label{HSBgeneral}
H_{{SR}} = \sum_{k}\frac{{p_k}^2}{2m_k}+\frac{m_k{\omega_k}^2}{2}\left(x_k-\frac{d_k}{{m_k\omega_k}^2}\sigma_z\right)^2,
\end{equation}
with $d_k$ denoting the coupling parameters for the interaction between the spin and the bosonic field.
In terms of creation and annihilation operators, Eq.~(\ref{HSBgeneral}) decomposes into a reservoir Hamiltonian and an interaction term. 
We define position $x^h_k = (\frac{1}{2m_k^h\omega^h_k})^{1/2}(b_k^{\dagger} + b_k)$ and momentum $p^h_k = i(\frac{m_k^h\omega^h_k}{2})^{1/2}(b_k^{\dagger} - b_k)$ for the hot reservoir $R_h$, and analogous relations for the cold reservoir $R_c$. The self-Hamiltonians for the reservoirs 
are then given by
\begin{eqnarray}
H_{R_h} & = & \sum_{k}\omega^h_{k}(b_{k}^{\dagger}b_{k}+{1}/{2}),\\
H_{R_c} & = & \sum_{q}\omega^c_{q}(c_{q}^{\dagger}c_{q}+{1}/{2}),
\end{eqnarray}
written in terms of creation (annihilation) operators $b_k^{\dagger}$ $(b_k)$ and $c_q^{\dagger}$ $(c_q)$ for the hot and cold reservoir, respectively. 
The corresponding oscillator frequencies are denoted by $\omega^h_k$ and $\omega^c_q$. 
The interactions between the TLS and each reservoir, with all constants subsumed into couplings $f_k^h \equiv \frac{d_k}{\sqrt{m^h_k \omega^h_k}}$ and $f_q^c \equiv \frac{d_q}{\sqrt{m^c_q \omega^c_q}}$, may then be written 
\begin{eqnarray}
H_{I_{h}} & = & -\sigma_{z}\sum_{k}f_{k}^h(b_{k}^{\dagger}+b_{k})+\sum_k{(f_k^h)}^2/\omega_k^h,\\
H_{I_{c}} & = & -\sigma_{z}\sum_{q}f_q^c(c_{q}^{\dagger}+c_{q})+\sum_q{(f_q^c)}^2/\omega_q^c.
\end{eqnarray}
The full Hamiltonian is given by the sum of all terms, 
\begin{equation}
H(t)=H_{S}(t)+H_{R_h}+H_{R_c}+H_{I_{h}}+H_{I_{c}},
\end{equation}
where it should be remembered that the interactions are only present along the relevant isochores.
In the following, we shall omit terms in the reservoir and interaction Hamiltonians proportional to the identity 
as they do not contribute 
when evaluating a complete engine cycle, nor help in defining sign conventions as was the case for $H_S(t)$.

\subsection{Reaction Coordinate Formalism}
Before turning to 
detailed calculations of the Otto cycle performance, 
we review elements of the reaction (or collective) coordinate formalism that are pertinent to the analysis of a heat engine. A complete and thorough derivation of the mapping is given in Refs~\cite{PhysRevA.90.032114, Iles-Smith2015}.

\begin{figure}[t]
\includegraphics[width=0.8\textwidth]{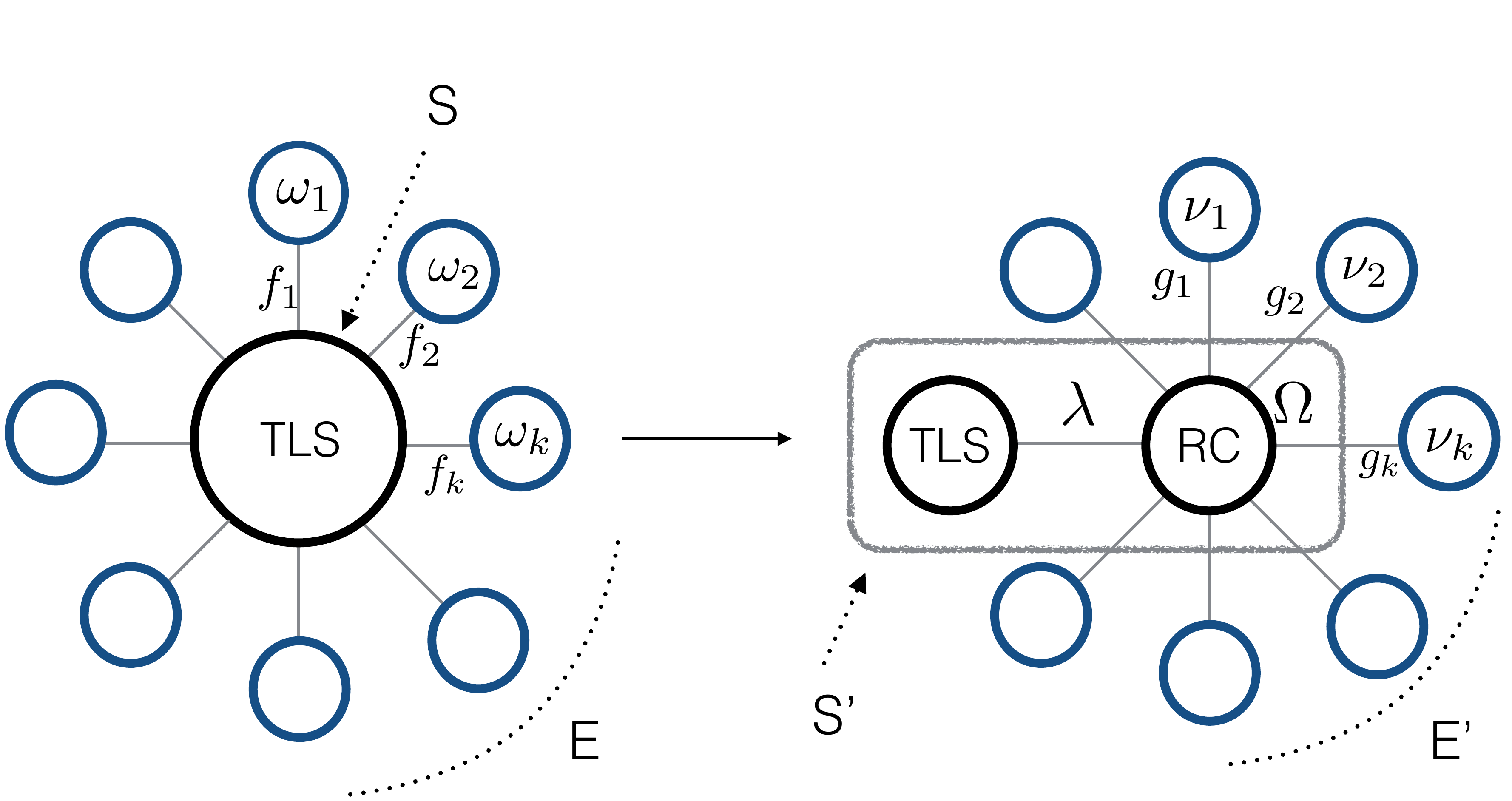}
\caption{Schematic of the RC mapping. Original picture (left): The TLS $S$ is strongly coupled via $f_k$ to an environment of harmonic oscillators $E$ with natural frequencies $\omega_k$. Mapped picture (right): An enlarged system $S'$---consisting of the TLS strongly coupled via $\lambda$ to a RC with natural frequency $\Omega$---is weakly coupled via $g_k$ to a residual oscillator environment $E'$ of natural frequencies $\nu_k$. \label{RCschematic}}
\end{figure}

The reaction-coordinate (RC) approach to strong coupling is based upon a unitary mapping of the system-reservoir Hamiltonian, schematically depicted in Fig.~\ref{RCschematic}. In our case, a TLS $S$ interacting with a multimode harmonic oscillator environment $E$ is mapped to 
an enlarged system $S'$, consisting of the TLS and a single collective degree of freedom of the environment called the RC, 
which interact strongly. The RC is also weakly coupled to a redefined residual environment $E'$, though this does not restrict the coupling strength between $S$ and $E$ in the original picture. 
For example, considering 
a single reservoir, the original Hamiltonian is given by (ignoring irrelevant terms in the reservoir and interaction Hamiltonians as stated)
\begin{eqnarray}
H & =& \frac{\mu(t)}{2}I + \frac{\epsilon(t)}{2}\sigma_{z}+\frac{\Delta(t)}{2}\sigma_{x} +\sum_{k}\omega_{k}b_{k}^{\dagger}b_{k}\nonumber\\
&& -\sigma_{z}\sum_{k}f_{k}(b_{k}^{\dagger}+b_{k}),
\label{Sb hamiltonian}
\end{eqnarray}
where we shall characterise the system-reservoir interaction by means of the spectral density 
\cite{RevModPhys.59.1}
\begin{align}
J(\omega)\equiv\sum_k f_k^2 \delta(\omega - \omega_k).
\label{SB spectral density}
\end{align}
In the cycle computations, we shall take the continuum limit for the bath oscillators, and assume the following functional form of the spectral density for each reservoir, 
\begin{align}
J(\omega)=\frac{\alpha\omega\omega_{c}}{\omega^{2}+\omega_{c}^{2}},
\label{SB spectral density function}
\end{align} 
where $\alpha$ is the 
coupling strength and $\omega_{c}$ is a cutoff frequency. 

The mapped Hamiltonian is then obtained by defining a collective coordinate (the RC) with creation  
operator $a^{\dagger}=(1/\lambda)\sum_kf_{k}b_{k}^{\dagger}$, which satisfies bosonic commutation relations for $\lambda=\sqrt{\sum_kf_k^2}$, such that
\begin{equation}\label{mappingHI}
H_I=-\sigma_{z}\sum_{k}f_{k}(b_{k}^{\dagger}+b_{k})=-\lambda\sigma_{z}(a^{\dagger}+a).
\end{equation} 
The reservoir Hamiltonian maps as 
\begin{align}\label{mappingHB}
H_R&=\sum_{k}\omega_{k}b_{k}^{\dagger}b_{k}\nonumber\\
&=\Omega a^{\dagger}a +\sum_{k}g_{k}(a^{\dagger}+a)(r_{k}^{\dagger}+r_{k})+\sum_{k}\nu_{k}r_{k}^{\dagger}r_{k},
\end{align}
while the system Hamiltonian remains unchanged. 
The full mapped Hamiltonian thus becomes~\footnote{The mapped Hamiltonian defined in Eq.~(\ref{RC hamiltonian}) differs from
  that given in  Eq.~$(2)$ of Ref.~\onlinecite{PhysRevA.90.032114} in that it does not include an extra term that is quadratic in the RC creation and annihilation operators, and the couplings $g_k$. As explained in Ref.~\cite{PhysRevA.90.032114}, there
  the term is added to cancel energy shifts that appear in the master equation governing the dynamics of the enlarged
  system. However, as we shall be concerned here with steady state solutions for the enlarged system, which are independent of the quadratic term, it has no bearing in our case, and we work directly with the mapped Hamiltonian as given in Eq.~(\ref{RC hamiltonian}).}
\begin{align}
\tilde{H}=& \frac{\mu(t)}{2}I + \frac{\epsilon(t)}{2}\sigma_{z}+\frac{\Delta(t)}{2}\sigma_{x}-\lambda\sigma_{z}(a^{\dagger}+a)\nonumber \\
&+\Omega a^{\dagger}a +\sum_{k}g_{k}(a^{\dagger}+a)(r_{k}^{\dagger}+r_{k})+\sum_{k}\nu_{k}r_{k}^{\dagger}r_{k}.
 \label{RC hamiltonian}
\end{align}
Here, the RC has natural frequency $\Omega$, it couples to the system with strength $\lambda$, and to 
the residual environment via $g_k$, whose oscillator excitations at natural frequency $\nu_{k}$ are created (annihilated) by $r_{k}^{\dagger}$ ($r_k$). These obey bosonic commutation relations and commute with the RC operators as they describe different modes in the mapped representation. Since the residual bath is traced out when deriving a master equation for the enlarged system $S'$, explicit expressions for the frequencies $\nu_k$ and the operators $r_k$ are not required. It suffices to find a functional form in the continuum limit for the spectral density function $\tilde{J}(\omega) \equiv \sum_k g_k^2 \delta(\nu - \nu_k) $ characterising the coupling between $S'$ and the residual bath $E'$, as well as the parameters $\Omega$ and $\lambda$, such that the Heisenberg equations of motion for the TLS are equivalent in both pictures. 
This results in
\begin{align}
\Omega = 2 \pi \gamma \omega_c, \label{param1} \\
\lambda = \sqrt{\frac{\pi \alpha \Omega}{2}},
\label{param2}
\end{align}
and 
\begin{align}
\tilde{J}(\omega)=\gamma \omega e^{-{\omega}/{\Lambda}},
\end{align}
where we set the (free) parameter $\gamma = \frac{\sqrt{\epsilon^2 + \Delta^2}}{2 \pi \omega_c}$ and eventually take to infinity the cutoff frequency $\Lambda$, in order to ensure that the original form of spin-boson spectral density is still accurately represented post mapping~\cite{PhysRevA.90.032114}.

A standard Born-Markov treatment~\cite{Breuer:2002aa} of the enlarged open quantum system $S'$ now leads to a second order master equation where the {\it strong} coupling between the TLS and the RC is treated exactly, while only the weakly coupled 
residual environment 
is traced out. 
The master equation can be solved numerically to obtain the dynamics of the TLS (or RC) as in Refs.~\cite{PhysRevA.90.032114,Iles-Smith2015}, 
where the number of RC basis states, labelled $n$ here, is truncated at a sufficient size to ensure convergence. Benchmarking against other numerical techniques has validated that this combination of Hamiltonian mapping and second order master equation is capable of very accurately capturing both the TLS dynamics and steady-states over 
a wide range of parameters~\cite{PhysRevA.90.032114,Iles-Smith2015}, particularly when non-Markovian and strong reservoir coupling effects preclude second order (i.e.~Born-Markov) expansions in the original unmapped representation. Furthermore, the mapping also allows access to properties of the reservoir through the RC itself, as well as to the system-reservoir correlations, something that is often impossible within other open systems 
approaches. This will in fact prove to be crucial in studying the heat engine at strong coupling in the following sections, as it 
allows a tractable analysis of the Otto cycle to be formulated in terms of the {\it full} system-reservoir Hamiltonian, including interactions and the resulting correlations.

Of particular importance in the present context is then the steady state solution of the master equation governing the dynamics of $S'$, as this will determine the state of the correlated system and reservoir at the end of each isochore. Since the coupling to the residual environment is treated according to the Born-Markov approximations within the RC formalism, this steady state is given by a thermal state of the {\emph{mapped}} system Hamiltonian
\begin{equation}
\tilde{H}_{S'}\equiv \frac{\mu(t)}{2}I + \frac{\epsilon(t)}{2}\sigma_{z}+\frac{\Delta(t)}{2}\sigma_{x}-\lambda\sigma_{z}(a^{\dagger}+a)+\Omega a^{\dagger}a,
\end{equation}
as
\begin{align}
\tilde{\rho}_{S'}=\frac{\exp\left(-\beta \tilde{H}_{S'}\right)}{\textrm{tr}\left[\exp\left(-\beta \tilde{H}_{S'}\right)\right]},
\label{noncanonical thermal state}
\end{align}
where $\beta=1/k_BT$ is the inverse temperature. The full state is then approximately
\begin{equation}\label{fullnoncanonicalstate}
\tilde{\rho}\approx\tilde{\rho}_{S'}\otimes\tilde{\rho}_{E'}
\end{equation}
where 
\begin{equation}\label{residualthermal}
\tilde{\rho}_{E'}=\frac{\exp(-\beta\tilde{H}_{E'})}{{\rm tr}[\exp(-\beta\tilde{H}_{E'})]},
\end{equation}
is a Gibbs thermal state of the residual environment with $\tilde{H}_{E'}=\sum_k\nu_kr_k^{\dagger}r_k$. 
One can obtain the reduced state of the TLS by performing a partial trace over the RC degrees of freedom, $\rho_S = \textrm{tr}_{RC+E'}[\tilde{\rho}]=\textrm{tr}_{RC}[\tilde{\rho}_{S'}]$, which in general does not take the form of a canonical Gibbs thermal state due to the correlations generated between the system and reservoir via their non-negligible interactions. Likewise, the reduced state of the RC 
can be obtained from a partial trace over the TLS. Eqs.~(\ref{noncanonical thermal state}) and (\ref{fullnoncanonicalstate}) are thus central to our analysis of the Otto cycle beyond weak coupling assumptions.

This can be exemplified 
by considering the energy expectation with respect to the full Hamiltonian
\begin{equation}\label{Hexpectation}
\langle H\rangle={\rm tr}[H\rho],
\end{equation}
where $H$ is given by Eq.~(\ref{Sb hamiltonian}) and $\rho={\exp(-\beta H})/{{\rm tr}[\exp(-\beta H)]}$ is a thermal state of the interacting system and reservoir in the original representation. 
Eq.~(\ref{Hexpectation}) is often difficult to evaluate---requiring advanced numerical techniques---without making a factorisation 
assumption between the system and reservoir. However, under the RC treatment it becomes
\begin{align}\label{Hexpecmapping}
\langle H\rangle&={\rm tr}[\tilde{H}\tilde{\rho}]\nonumber\\
&\approx{\rm tr}[\tilde{H}_{S'}\tilde{\rho}_{S'}]+{\rm tr}[\tilde{H}_{E'}\tilde{\rho}_{E'}],
\end{align}
where in the last line we have used Eq.~(\ref{fullnoncanonicalstate}) for the mapped density operator and the fact that ${\rm tr}[\sum_kg_k(a^{\dagger}+a)(r_k^{\dagger}+r_k)\tilde{\rho}_{E'}]=0$. Thus, within the RC approach, the average energy of the interacting system and reservoir reduces to a sum of thermal expectations for the enlarged mapped system Hamiltonian and the residual bath, 
substantially simplifying calculations in the strong coupling regime. 
This is also intuitively appealing as it draws a natural boundary between the system and reservoir 
at finite coupling strength to link to standard thermodynamics, i.e.~the residual environment provides a well defined temperature as well as a reference for energy absorption and dissipation even at strong coupling.

\subsection{Generalised Otto Cycle Analysis \label{weak}}

We shall now present a detailed 
analysis of the Otto cycle beyond weak system-reservoir coupling and vanishing correlations. We accomplish this by considering energetic changes with respect to the full system-reservoir Hamiltonian $H$ and state $\chi$, rather than just the internal system Hamiltonian $H_S$ and the reduced system state $\rho_S$. This leads to expressions that 
may be evaluated for arbitrary interaction strength using the RC formalism, 
and further 
reduce to the standard weak coupling forms under the assumption of 
fully factorising system-reservoir states.

\subsubsection{Hot isochore}

Without loss of generality, we consider starting the analysis of the cycle at point $A'$ (see Fig.~\ref{quantum-otto-cycle}), 
when the interaction between the system and the hot reservoir has just been switched on (assumed instantaneous). The interaction with the cold reservoir is not present, however. The Hamiltonian along the isochore is given by
\begin{equation}
H=H_{S}^{A'}+H_{R_h}+H_{R_c}+H_{I_{h}},
\end{equation}
where 
\begin{equation}\label{HSAPHSB}
H_S^{A'}=H_S^B=\frac{\mu_h}{2}I+\frac{\epsilon_h}{2}\sigma_z+\frac{\Delta_h}{2}\sigma_x,
\end{equation}
is unchanging along the stroke, with $H_S^{A'}$ and $H_S^B$ representing the system Hamiltonian at point $A'$ and $B$, respectively. 
At the end of the stroke (point $B$) the full state of the system and both reservoirs has relaxed to equilibrium, which we write as
\begin{equation}
\chi^{B}=\rho_{h} \otimes \rho_{R_c},
\end{equation}
where 
\begin{equation} \label{rhoh}
\rho_{h}=\frac{\exp\left[-\beta_{h}\left(H_{S}^{B}+H_{R_h}+H_{I_{h}}\right)\right]}{\textrm{tr}\left\{ \exp\left[-\beta_{h}\left(H_{S}^{B}+H_{R_h}+H_{I_{h}}\right)\right]\right\} },
\end{equation}
is the equilibrium state of the interacting TLS and hot reservoir and $\rho_{R_c}$ 
represents 
the state of the uncoupled cold reservoir. 
We have made the assumption that there are no correlations between the system and the cold reservoir at this stage in the cycle, since they are non-interacting, and hence the state $\rho_{R_c}$ may be factored out. 

In general, 
$\rho_h$ 
is difficult to evaluate for a non-vanishing interaction term $H_{I_h}$. However, by employing the RC formalism, in particular Eq.~(\ref{fullnoncanonicalstate}), we may write 
\begin{equation}\label{rhoH}
\rho_h=\frac{\exp(-\beta_h\tilde{H}_{S'}^B)}{{\rm tr}[\exp(-\beta_h\tilde{H}_{S'}^B)]}\otimes\tilde{\rho}_{E'_h}=\tilde{\rho}_{S'_h}\otimes\tilde{\rho}_{E'_h},
\end{equation}
which is far easier to calculate for finite system-reservoir coupling.  
Here $\tilde{\rho}_{E'_h}$ is a thermal state of the hot reservoir residual environment (self Hamiltonian $\tilde{H}_{E'_h}$) at inverse temperature $\beta_h=1/k_BT_h$ 
and
\begin{equation}
\tilde{H}_{S'}^B=H_S^B-\lambda_h\sigma_z(a_h^{\dagger}+a_h)+\Omega_ha_h^{\dagger}a_h,
\end{equation}
is the mapped (enlarged) system Hamiltonian including a RC for the hot reservoir.

For the cycle analysis, we are interested in the average energy at the end of the stroke, point $B$. This is obtained by taking the trace of the Hamiltonian with the full state
\begin{align}\label{energyB}
\langle H\rangle^{B} & =  \textrm{tr}\left[\left(H_{S}^{B}+H_{R_h}+H_{R_c}+H_{I_{h}}\right)\rho_{h}\rho_{R_c}\right]\nonumber\\
 & =  \textrm{tr}\left[\tilde{H}_{S'}^{B}\tilde{\rho}_{S'_h}\right]+\textrm{tr}\left[\tilde{H}_{E'_h}\tilde{\rho}_{{E'_h}}\right]+\textrm{tr}\left[H_{R_c}\rho_{R_c}\right],
 \end{align}
where in the second line we have re-written the trace using the RC approach, as in Eq.~(\ref{Hexpecmapping}). This distinguishes correlated and uncorrelated contributions due to the system and hot reservoir, the latter of which depends only on the residual thermal bath and will cancel out in the full cycle analysis. The presence of the residual thermal bath is still important, however, as it provides a reference against which we can define energy absorption along the hot isochore. 
The third term in Eq.~(\ref{energyB}) is the internal energy of the cold reservoir. 
To remain close to the spirit of the classical Otto cycle, we shall assume that 
the cold reservoir thermalises when uncoupled from the system, and likewise with the hot reservoir. 
In this way, whenever the coupling between the system and either reservoir is switched on, the reservoir is always initially in a thermal equilibrium state of well defined temperature, with correlations then generated along the subsequent isochore. 
It follows then that $\rho_{R_c}$ should be taken at this point to be a Gibbs thermal state at inverse temperature $\beta_c=1/k_BT_c$: 
\begin{equation}\label{thermalc}
\rho_{R_c}=\rho_{th_{c}}=\frac{\exp(-\beta_cH_{R_c})}{{\rm tr}[\exp(-\beta_cH_{R_c})]}.
\end{equation}

In the standard weak coupling analysis, one makes the assumption that the full state remains separable at all times with 
each reservoir 
in a thermal state at its given temperature. Under these conditions  
the state $\rho_h$ factorises into a product of the system state $\rho_{S_h}$ and the hot reservoir state $\rho_{th_h}$. Each state is of canonical Gibbs form with respect to the relevant self-Hamiltonian, which for the system reads 
\begin{align}\label{rhoSh}
\rho_{S_h} =  \frac{\exp\left(-\beta_{h}H_{S}^B\right)}{\textrm{tr}\left[\exp\left(-\beta_{h}H_{S}^B\right)\right]},
\end{align}
with $H_S^B$ defined in Eq.~(\ref{HSAPHSB}), and for the hot reservoir is 
\begin{equation}\label{thermalh}
\rho_{th_h}=\frac{\exp(-\beta_hH_{R_h})}{{\rm tr}[\exp(-\beta_hH_{R_h})]}.
\end{equation} 
With the system Hamiltonian defined in equations (\ref{Hs}) and (\ref{mu}), the expression for the average energy then reduces to
\begin{eqnarray}
\langle H\rangle^{B}_{weak} & = & \frac{\mu_{h}}{2}\left[1-\tanh\left(\frac{\mu_{h}}{2k_{B}T_{h}}\right)\right]+\langle H_{R_h}\rangle_{th}\nonumber\\
 &&\:{+}\langle H_{R_c}\rangle_{th},
\end{eqnarray}
where we denote the reservoir thermal expectations by $\langle H_{R_i}\rangle_{th}={\rm tr}[H_{R_i}\rho_{th_i}]$, 
for $i=h,c$.

The interaction between the TLS and the hot reservoir is then switched off. In order to explicitly consider any cost associated with this step, we define the point after decoupling as $B'$ and denote the corresponding Hamiltonian as 
\begin{equation}
H^{B'}=H^B_{S}+H_{R_h}+H_{R_c},
\end{equation}
with energy
\begin{equation}
\langle H\rangle^{B'} = \textrm{tr}\left[\left(H_{S}^{B}+H_{R_h}+H_{R_c}\right)\chi^{B'}\right].
\end{equation}
If $H_{I_h}$ is switched off instantaneously, then the full state has no time to change between points $B$ and $B'$ and so $\chi^{B'} = \chi^{B}=\rho_h \otimes \rho_{th_c}$, and 
the work in decoupling 
is given by 
\begin{equation} \label{Hi}
\langle H\rangle^{B'}-\langle H\rangle^{B}=-\textrm{tr}\left[H_{I_{h}}\rho_{h}\right].
\end{equation}
For the standard weak coupling treatment $\rho_h = \rho_{S_h} \otimes \rho_{th_h}$ and the energy cost equation (\ref{Hi}) associated with decoupling from the environment evaluates identically to zero. 
%if the 
%hot reservoir 
%is in a thermal state that factorises from the system, 
Within the RC approach for finite coupling on the other hand, the state is given by equation (\ref{rhoH}) and we obtain
\begin{equation}
\langle H\rangle^{B'}-\langle H\rangle^{B}=\lambda_h\textrm{tr}\left[\sigma_z(a_h^{\dagger}+a_h)\tilde{\rho}_{S'_h}\right],
\end{equation}
which is generally non-zero, and the work contribution due to decoupling must therefore be included in the cycle analysis. 
If this contribution were negative,
then we would extract work. However, we shall see that in the examples considered here it is positive, and thus represents a cost. 
As an alternative to attempt to mitigate this cost we also consider switching the interaction off adiabatically, the details of which are described in the Appendix.

\subsubsection{Isentropic expansion}

With the system now interacting with neither of the two reservoirs, the parameters in $H_S$ are changed such that  $\mu_{h}\rightarrow\mu_{c}$,
$\epsilon_{h}\rightarrow\epsilon_{c}$, $\Delta_{h}\rightarrow\Delta_{c}$. The TLS Hamiltonian at the end of the stroke (point C) becomes 
\begin{equation}
H_{S}^{C}=\frac{\mu_{c}}{2}I+\frac{\epsilon_{c}}{2}\sigma_{z}+\frac{\Delta_{c}}{2}\sigma_{x}.
\end{equation}
In the usual treatment of the Otto cycle, it is assumed that these parameters are changed slowly enough such that the quantum adiabatic theorem holds.
We are then interested in the average energy along the stroke
\begin{equation}
\langle H(t)\rangle=\textrm{tr}\left[\left(H_{S}(t)+H_{R_h}+H_{R_c}\right)\chi(t)\right].
\label{Wext}
\end{equation}
To see how the adiabatic theorem applies, one can define a unitary transformation %Defining a unitary operator 
$V\left(t\right)=\exp\left[-i\sigma_{y}{\theta\left(t\right)}/{2}\right]$, 
with
$\theta\left(t\right)=\tan^{-1}\left[{\Delta\left(t\right)}/{\epsilon\left(t\right)}\right]$, and consider the Hamiltonian $H'$ in the frame defined by $V(t)$. One finds
\begin{eqnarray}
H'(t) & = & V^{\dagger}(t) H(t) V(t) + i  \dot{V}^{\dagger}(t)V(t)\nonumber\\
& = & \frac{\mu(t)}{2}\left( I + \sigma_z\right) - \frac{\dot{\theta} (t)}{2} \sigma_y +H_{R_h}+H_{R_c}.
\end{eqnarray}
In the adiabatic limit, one assumes $\dot{\theta}(t) \ll 1$ and the transformation approximately diagonalises the system Hamiltonian $H_S(t)$. 
The average energy at any point along the stroke, given by equation (\ref{Wext}), may then be calculated in this new basis: 
\begin{align}
\langle H(t)\rangle = \frac{\mu\left(t\right)}{2}\left\{1+\textrm{tr}\left[\sigma_{z}\chi'(t)\right]\right\}+{\rm tr}\left[(H_{R_h}+H_{R_c})\chi(t)\right],
\end{align}
where $\chi'(t)\equiv V^{\dagger}(t)\chi(t)V(t)$.
The term 
$\textrm{tr}\left[\sigma_{z}\chi'(t)\right]$
remains constant. To see this, we may substitute in for $\chi'(t)\equiv U'(t)\chi'(0)U'^{\dagger}(t)$, 
where we define a time evolution operator in the transformed frame as $U'(t)\equiv T\exp\left[-i\int_0^t H_S'(\tau)d\tau\right]$, with $T$ denoting the time-ordered exponential. Thus, we have
\begin{eqnarray}
\text{tr}\left[\sigma_{z}\chi'(t)\right]&=&\text{tr}\left[\sigma_{z}U'(t)\chi'(0)U'^{\dagger}(t)\right]\nonumber\\
&=&\text{tr}\left[U'^{\dagger}(t)\sigma_{z}U'(t)\chi'(0)\right],
\end{eqnarray}
where we have used the cyclic property of the trace. In the adiabatic limit ($\dot{\theta}\ll1$), $H'_S(t)\approx(\mu(t)/2)(I+\sigma_z)$, such that the time evolution operator commutes with $\sigma_z$, and so $U'^{\dagger}(t)\sigma_{z}U'(t)=\sigma_z$.
This leaves us with
\[
\text{tr}\left[\sigma_{z}\chi'(t)\right]=\text{tr}\left[\sigma_{z}\chi'(0)\right],
\]
with the result that this term remains constant along the stroke.
The average energy along the stroke is then
\begin{align}
\langle H(t)\rangle &= \frac{\mu(t)}{2}\left\{1+\textrm{tr}\left[\sigma_{z}\chi(0)\right]\right\}+{\rm tr}\left[(H_{R_h}+H_{R_c})\chi'(t)\right]\nonumber\\
 &= \frac{\mu(t)}{\mu(0)}\langle H_{S}(0)\rangle +{\rm tr}\left[(H_{R_h}+H_{R_c})\chi(t)\right].
 \end{align}
As the reservoirs are evolving only under their own self-Hamiltonians, their energy expectations are unchanging. Hence, at the end of the stroke the energy of the system reads 
\begin{align}
\langle H\rangle^{C} & =  \textrm{tr}\left[\left(H_{S}^{C}+H_{R_h}+H_{R_c}\right)\chi^{C}\right]\nonumber\\
 & =  \frac{\mu_{c}}{\mu_{h}}\textrm{tr}\left[H_{S}^{B}\rho_{h}\right]+\text{tr}\left[H_{R_h}\rho_{h}\right]+\langle H_{R_c}\rangle_{th}.
\end{align}
The work extracted along the stroke is given by the difference in energy between the start and end points 
\begin{equation}
\langle H\rangle^{C}-\langle H\rangle^{B'}=\left(\frac{\mu_{c}}{\mu_{h}}-1\right)\textrm{tr}\left[H_{S}^{B}\rho_{h}\right],
\label{workoutstrong}
\end{equation}
which reduces in the standard weak coupling limit to
\begin{equation}
\langle H\rangle^{C}_{weak}-\langle H\rangle^{B'}_{weak}=\left(\frac{\mu_{c}}{2}-\frac{\mu_{h}}{2}\right)\left[1-\tanh\left(\frac{{\mu}_{h}}{2k_{B}T_{h}}\right)\right],
\label{workoutweak}
\end{equation}
where $\mu_h > \mu_c$. 

The interaction between the system and the cold reservoir is now switched on instantaneously, leading to point $C'$, 
where the energy is
\begin{equation}
\langle H\rangle^{C'} = \textrm{tr}\left[\left(H_{S}^{C}+H_{R_h}+H_{R_c}+H_{I_{c}}\right)\chi^{C'}\right],
\label{HC'}
\end{equation}
and $\chi^{C'}=\chi^{C}$.
We make the assumption that the hot and cold reservoirs rapidly relax to equilibrium at temperatures $T_h$ and $T_c$ respectively, when disconnected from the system. This means that when the interaction is turned on, the system is connecting to a reservoir in a thermal equilibrium state. Hence, the work associated with coupling to the cold reservoir, $\langle H\rangle^{C'}-\langle H\rangle^{C}$, evaluates to zero in both the weak coupling and RC treatments as the cold reservoir state $\rho_{R_c}$ is a thermal state at this point. Note that we do not consider the option of adiabatically turning on the coupling to the cold reservoir as this would negate the need for the cold isochore that follows, and thus has no equivalent in the standard Otto cycle. 

\subsubsection{Cold isochore}

The system is now allowed to reach equilibrium with the cold reservoir. In analogy to the hot isochore, the full state at the end of the stroke (point $D$) is given by 
\begin{equation}
\chi^{D}=\rho_{c} \otimes \rho_{th_h},
\end{equation}
where
\begin{equation}
\rho_{c}=\frac{\exp\left[-\beta_{c}\left(H_{S}^{D}+H_{R_c}+H_{I_{c}}\right)\right]}{\textrm{tr}\left\{ \exp\left[-\beta_{c}\left(H_{S}^{D}+H_{R_c}+H_{I_{c}}\right)\right]\right\} },
\label{coldthermstate}
\end{equation}
is a thermal state of the interacting TLS and cold reservoir and we assume that the hot reservoir has rethermalised along the stroke, see Eq.~(\ref{thermalh}). 

Employing the RC treatment to account for the interacting system and cold reservoir, the energy at point $D$ is written 
\begin{align}
\langle H\rangle^{D} & =  \textrm{tr}\left[\left(H_{S}^{C}+H_{R_h}+H_{R_c}+H_{I_{c}}\right)\rho_{c}\rho_{th_h}\right]\nonumber\\
 & =  \textrm{tr}\left[\tilde{H}_{S'}^{C}\tilde{\rho}_{S'_c}\right]
 +\textrm{tr}\left[\tilde{H}_{E'_{c}}\tilde{\rho}_{E'_c}\right]+\langle H_{R_h}\rangle_{th},
 \label{HD}
 \end{align}
with 
\begin{equation}\label{rhoC}
\rho_c=\frac{\exp(-\beta_c\tilde{H}_{S'}^C)}{{\rm tr}[\exp(-\beta_c\tilde{H}_{S'}^C)]}\otimes\tilde{\rho}_{E'_c}=\tilde{\rho}_{S'_c}\otimes\tilde{\rho}_{E'_c},
\end{equation}
where
\begin{equation}
\tilde{H}_{S'}^C=H_S^C-\lambda_c\sigma_z(a_c^{\dagger}+a_c)+\Omega_ca_c^{\dagger}a_c,
\end{equation}
is the RC mapped Hamiltonian for the system and cold reservoir, and $\tilde{\rho}_{E'_c}$ is a thermal state of the cold reservoir residual environment (self Hamiltonian $\tilde{H}_{E'_c}$) at inverse temperature $\beta_c$. In the weak coupling treatment, the state again reduces to the product of three thermal states for the system and the two reservoirs, 
such that
\begin{eqnarray}
\langle H\rangle^{D}_{weak} & = &  \frac{\mu_{c}}{2}\left[1-\tanh\left(\frac{\mu_{c}}{2k_{B}T_{c}}\right)\right]+\langle H_{R_h}\rangle_{th}\nonumber\\
 &&\:{+}\langle H_{R_c}\rangle_{th}.
\end{eqnarray}
In both the RC and weak coupling treatments, the energy exchanged with the cold reservoir is given by the energetic change across the isochore, 
$Q^{C'D}=\langle H\rangle^{D}-\langle H\rangle^{C'}$.

The interaction between the system and the cold reservoir is now switched off and we consider point $D'$, where the Hamiltonian contains no interaction terms, such that
\begin{equation}
\langle H\rangle^{D'} = \textrm{tr}\left[\left(H_{S}^{C}+H_{R_h}+H_{R_c}\right)\chi^{D'}\right].
\end{equation}
If the interaction is switched off instantaneously then $\chi^{D'}=\chi^{D}=\rho_c \otimes \rho_{th_h}$, and 
the work cost 
is therefore 
\begin{equation}
\langle H\rangle^{D'}-\langle H\rangle^{D}=-\textrm{tr}\left[H_{I_{c}}\rho_{c}\right].
\end{equation}
In the weak coupling treatment, the cold reservoir remains in a thermal state and factorises out of the expression for $\rho_c$, so that the cost again evaluates to zero. At strong coupling we obtain 
\begin{equation}
\langle H\rangle^{D'}-\langle H\rangle^{D}=\lambda_c\textrm{tr}\left[\sigma_z(a_c^{\dagger}+a_c)\tilde{\rho}_{S'_c}\right],
\end{equation}
within the RC approach, which is generally non-zero once more. As in the hot isochore, we consider the alternative of adiabatic decoupling in the Appendix. 

\subsubsection{Isentropic compression}

This stroke is analogous to the expansion stroke, with the system parameters changed back to their original values such that at point $A$ 
\begin{equation}
H_{S}^{A}=\frac{\mu_{h}}{2}I+\frac{\epsilon_{h}}{2}\sigma_{z}+\frac{\Delta_{h}}{2}\sigma_{x}.
\end{equation}
In the adiabatic limit the energy at point $A$ is given by
\begin{align}
\langle H\rangle^{A} & =  \textrm{tr}\left[\left(H_{S}^{A}+H_{R_h}+H_{R_c}\right)\chi^{A}\right]\nonumber\\
 & =  \frac{\mu_{h}}{\mu_{c}}\textrm{tr}\left[H_{S}^{C}\rho_{c}\right]+\textrm{tr}\left[H_{R_c}\rho_{c}\right]+\langle H_{R_h}\rangle_{th}.
\end{align}
The energy difference across the stroke, equivalent to the work done on the system, is
\begin{equation}
\langle H\rangle^{A}-\langle H\rangle^{D'}=\left(\frac{\mu_{h}}{\mu_{c}}-1\right)\textrm{tr}\left[H_{S}^{C}\rho_{c}\right],
\end{equation}
which reduces to
\begin{equation}
\langle H\rangle_{weak}^{A}-\langle H\rangle_{weak}^{D'}=\left(\frac{\mu_{h}}{2}-\frac{\mu_{c}}{2}\right)\left[1-\tanh\left(\frac{\mu_{c}}{2k_{B}T_{c}}\right)\right],
\end{equation}
within the weak coupling treatment. 
The coupling to the hot reservoir is now switched on instantaneously and we return to point $A'$ where the Hamiltonian contains the term $H_{I_h}$. The energy is thus given by
\begin{align}
\langle H\rangle^{A'} = & \textrm{tr}\left[(H_{S}^{B}+H_{R_h}+H_{R_c}+H_{I_{h}})\chi^{A'}\right]\nonumber\\
  = & \frac{\mu_{h}}{\mu_{c}}\textrm{tr}\left[H_{S}^{C}\rho_{c}\right]+\text{tr}\left[H_{R_c}\rho_{c}\right]+
  \langle H_{R_h}\rangle_{th}\nonumber\\
  = & \langle H\rangle^{A}, 
\end{align}
where, as in the cold reservoir case, there is no cost associated with switching on the coupling to the hot reservoir.

\begin{figure}[t]
(a)
\includegraphics[width=0.98\textwidth]{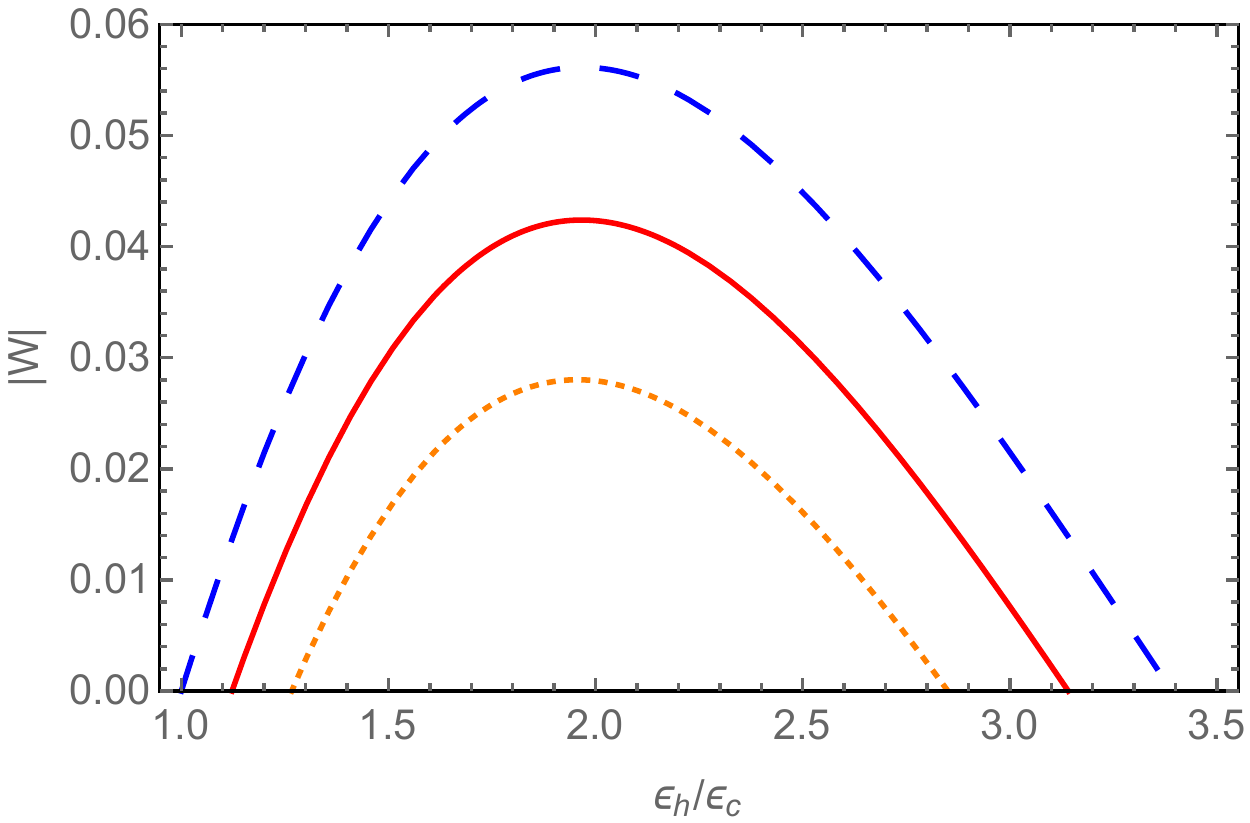}
(b)
\includegraphics[width=0.98\textwidth]{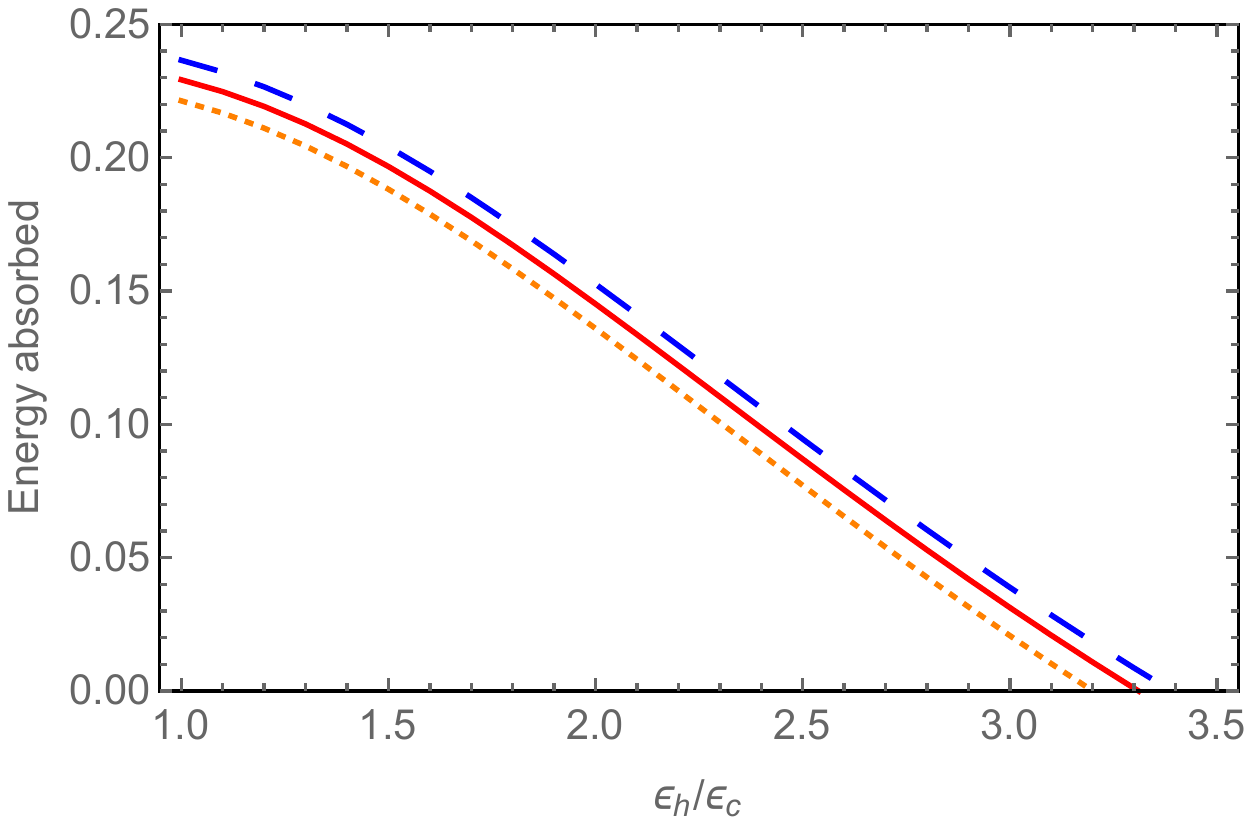}
\caption{{\bf Adiabatic limit.} Work output (a) and energy absorbed from the hot reservoir (b) for a quantum Otto cycle plotted as a function of the TLS bias at point $A$, $\epsilon_h$. {\it Blue dashed curves}: weak coupling; {\it Orange dotted curves}: strong coupling with instantaneous decoupling of the reservoirs; {\it Red solid curves}: strong coupling with adiabatic decoupling of the reservoirs. Parameters (in units of $\epsilon_c$): $\Delta_h = \Delta_c=1$, $\beta_{h}=1$, $\beta_{c}=2.5$, $\omega_{c}=2$, $\alpha=0.005$, and $n=30$ states are taken in the RC calculations.
\label{Workfigad}}
\end{figure}

\subsection{Work output and efficiency at strong coupling}

We are now in a position to evaluate the net work output and energy conversion efficiency of the strong coupling Otto cycle. 
The work output is obtained by summing 
the energetic changes 
along each isentropic stroke, including the work costs associated with decoupling:
\begin{align}
W  =&  W^{BB'}+W^{B'C}+W^{CC'}+W^{DD'}+W^{D'A}+W^{AA'}\nonumber\\
  =&  \left(\frac{\mu_{c}}{\mu_{h}}-1\right)\textrm{tr}\left[H_{S}^{B}\rho_{h}\right]+\left(\frac{\mu_{h}}{\mu_{c}}-1\right)\textrm{tr}\left[H_{S}^{C}\rho_{c}\right]\nonumber\\
 &   -\textrm{tr}\left[H_{I_{h}}\rho_{h}\right]-\textrm{tr}\left[H_{I_{c}}\rho_{c}\right].
 \label{Work}
\end{align}
Similarly, the energy transferred into the system
is given by the energetic change 
along the hot isochore: 
\begin{align}
Q^{A'B}  = &  \langle H\rangle^{B}-\langle H\rangle^{A'}\nonumber\\
 = & \textrm{tr}\left[H_{S}^{B}\rho_{h}\right]-\frac{\mu_{h}}{\mu_{c}}\textrm{tr}\left[H_{S}^{C}\rho_{c}\right]\nonumber\\
 &+\textrm{tr}\left[H_{R_h}\left(\rho_{h}-\rho_{th_h}\right)\right]
+\textrm{tr}\left[H_{R_c}\left(\rho_{th_c}-\rho_{c}\right)\right]\nonumber\\
&+\textrm{tr}\left[H_{I_{h}}\rho_{h}\right].
\label{Heat}
\end{align}

In the weak coupling treatment, the reservoirs remain in thermal equilibrium such that their internal energies 
cancel out and the interaction terms evaluate to zero. The expressions for the net work output and energy flow from the hot reservoir then reduce to the standard forms
\begin{align}\label{workweak}
W_{weak} = &\langle H\rangle^{C'}-\langle H\rangle^{B}+\langle H\rangle^{A'}-\langle H\rangle^{D}\nonumber\\
 = & \frac{1}{2}\left(\mu_{c}-\mu_{h}\right)\left[\tanh\left(\frac{\mu_{h}}{2k_{B}T_{h}}\right)-\tanh\left(\frac{\mu_{c}}{2k_{B}T_{c}}\right)\right]
\end{align}
and
\begin{equation}\label{heatweak}
Q_{weak}   = \frac{\mu_{h}}{2}\left[\tanh\left(\frac{\mu_{c}}{2k_{B}T_{c}}\right)-\tanh\left(\frac{\mu_{h}}{2k_{B}T_{h}}\right)\right],
\end{equation}
respectively. 

The efficiency of the engine is given as usual by the ratio of the net work output to the energy absorbed by the system along the hot isochore, 
\begin{equation}
\eta = \frac{W}{Q^{A'B}}. 
\end{equation}
For weak coupling this reduces to 
\begin{equation}\label{efficiencyweak}
\eta_{weak} = 1 - \frac{{\mu_c}}{{\mu_h}},
\end{equation}
as expected.

\section{Otto cycle results \label{results}}

\begin{figure}[t]
\includegraphics[width=0.98\textwidth]{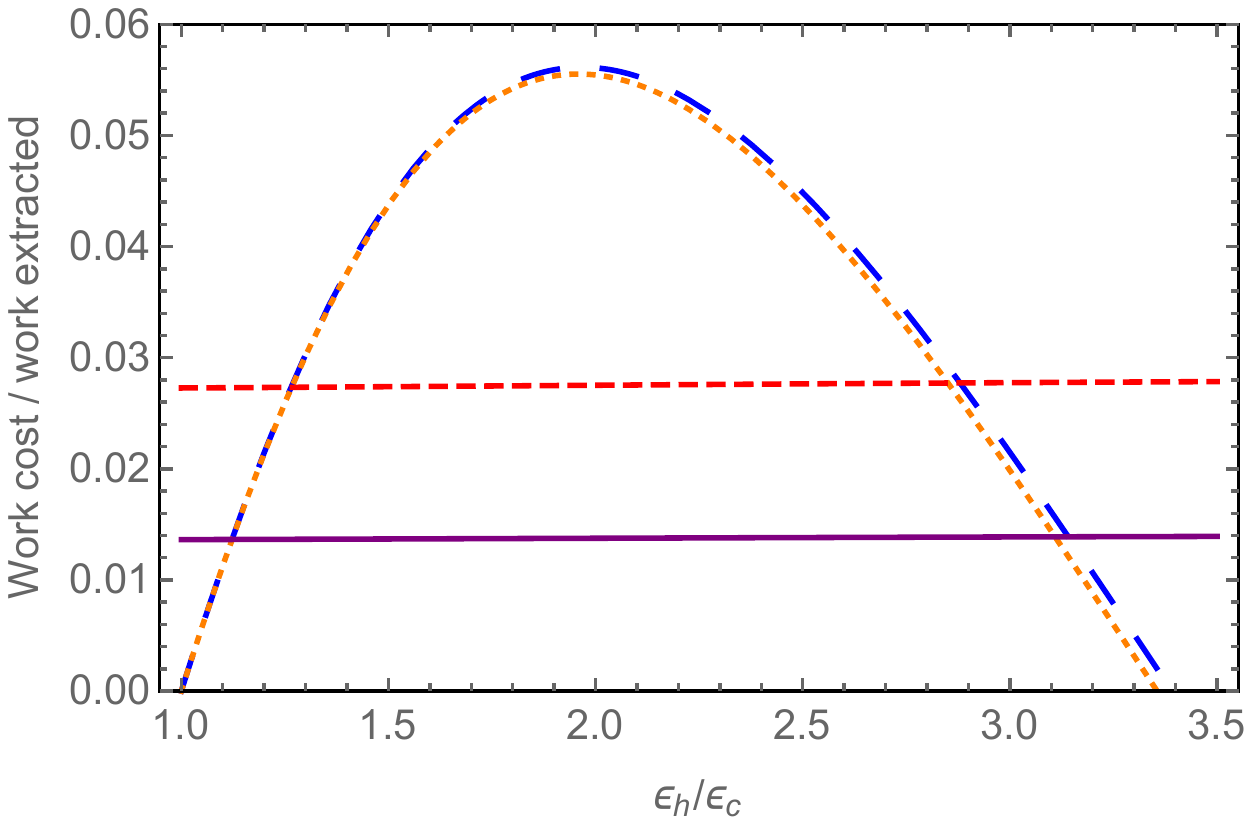}
\caption{{\bf Adiabatic limit.} Net work output of the Otto cycle split into decoupling work cost and magnitude of the remaining work extracted, 
plotted as a function of the TLS bias at point $A$, $\epsilon_h$. \textit{Blue dashed curve:} weak coupling (no decoupling cost) and strong coupling work extracted ignoring work cost (adiabatic decoupling case); \textit{Orange dotted curve:} strong coupling work extracted ignoring work cost (instantaneous decoupling case); \textit{Red dashed line:} strong coupling cost of instantaneous decoupling; \textit{Purple solid line:} strong coupling cost of adiabatic decoupling. Parameters as in Fig.~\ref{Workfigad}.
\label{WorkOutad}}
\end{figure}

We shall now present some specific examples to explore the impact of strong coupling and system-reservoir correlations on the quantum Otto engine's 
performance. We begin by considering adiabatic isentropic strokes, as outlined above, though for completeness we shall subsequently extend the analysis to the opposite limit of sudden isentropes as well. In both instances, we shall see through comparison to the weak coupling analysis that strong coupling acts to reduce the performance of the cycle. This is also consistent with findings 
for continuous heat engines in the strong coupling regime~\cite{doi:10.1021/acs.jpclett.5b01404}. However, in our case, we can  identify 
the work costs imposed in 
decoupling the system and reservoirs 
after each isochore as the primary reason for the reduced performance, which is thus distinct from considerations for continuous engines. In fact, as we shall see, in the absence of this cost, it is possible for the strong coupling engine to output more work than its weak coupling counterpart, though in all cases we have explored the decoupling cost outweighs this benefit. 

%\begin{figure}
%\vspace{0.1cm}
%\includegraphics[width=0.97\textwidth]{adiabatic_efficiency.pdf}
%\caption{{\bf Adiabatic limit.} Efficiency of the quantum Otto engine plotted as a function of the TLS bias at point $A$, $\epsilon_h$. \textit{Blue dashed curve:} weak coupling; \textit{Orange dotted curve:} strong coupling with instantaneous decoupling of the reservoirs; {\it Red solid curve:} strong coupling with adiabatic decoupling of the reservoirs. 
%Parameters as in Fig.~\ref{Workfigad}. 
%\label{efficiencyfigad}}
%\end{figure}

\begin{figure}
\vspace{0.1cm}
\includegraphics[width=0.97\textwidth]{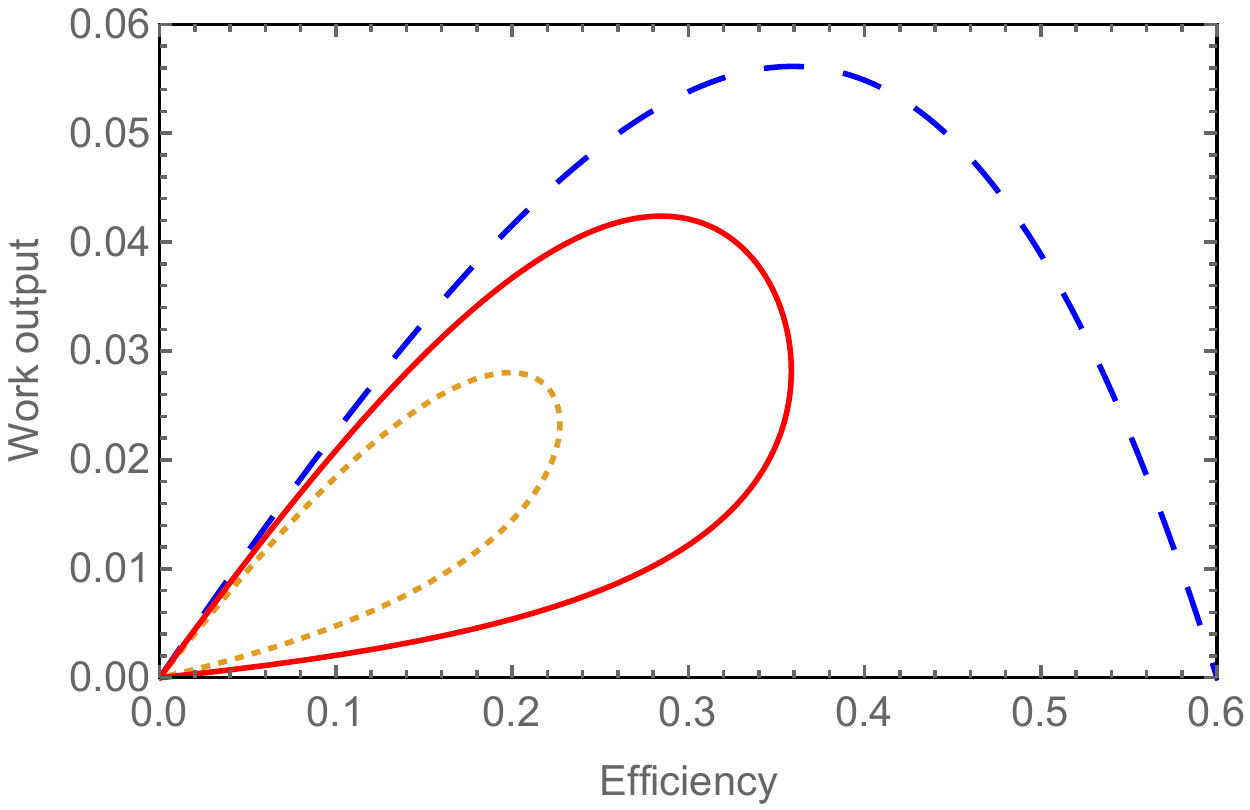}
\caption{{\bf Adiabatic limit.} Parametric plots of work output against efficiency in the adiabatic regime for a quantum Otto cycle plotted by varying the TLS bias at point $A$, $\epsilon_h$. {\it Blue dashed curve}: weak coupling; {\it Red solid curve}: strong coupling with adiabatic decoupling of the reservoirs; {\it Orange dotted curve}: Strong coupling with instantaneous decoupling of the reservoirs. Parameters (in units of $\epsilon_c$): $\Delta_h = \Delta_c=1$, $\beta_{h}=1$, $\beta_{c}=2.5$, $\omega_{c}=2$, $\alpha=0.005$, and $n=30$ states are taken in the RC calculations. 
\label{efficiencyfigad}}
\end{figure}

\subsection{Adiabatic isentropic strokes}

As considered in Section~\ref{weak}, in the adiabatic limit the isentropic strokes of the cycle are carried out slowly enough for the quantum adiabatic theorem to hold. 
In Fig.~\ref{Workfigad} we show representative plots of the work output and energy absorbed from the hot reservoir in this case, 
here as a function of the TLS bias at point $A$. 
We compare the weak coupling limit 
(dashed curves), 
Eqs.~(\ref{workweak}) and~(\ref{heatweak}), with our two alternative cycles in the strong coupling regime, which consider either instantaneous (dotted curves) or adiabatic (solid curves) decoupling of the reservoirs from the system. 
For strong coupling, all calculations have been performed using the RC formalism, as previously described.  

In the weak coupling limit, net work output vanishes when ${\mu_h}/{\mu_c}=1$, since the engine requires non-zero work input in order to operate. Similarly, from Eq.~(\ref{workweak}) it can be seen that for $W_{weak}$ to be negative (representing work output) the condition 
${\mu_h}/{\mu_c} < {T_h}/{T_c}$ must be satisfied, meaning that at ${\mu_h}/{\mu_c}={T_h}/{T_c}$ the work output vanishes as well, as does the energy absorbed from the hot reservoir. We note that  
beyond this point for weak coupling, ${\mu_h}/{\mu_c} > {T_h}/{T_c}$, the engine turns over to operate 
instead as a refrigerator, absorbing energy from the cold reservoir and dissipating into the hot reservoir.

It is clear that the strong coupling treatments yield lower work outputs 
than the weak coupling calculations, and that work 
cannot be extracted right up to the 
limit of ${\mu_h}/{\mu_c}={T_h}/{T_c}$. 
They do, however, show a reduction in the energy absorbed from the hot reservoir. Note also that in contrast to the weak coupling case, for strong coupling the work output and energy absorbed do not change sign at the same point, meaning that there are regimes in which the strong coupling cycle acts neither as an engine nor as a refrigerator. 
Looking more closely at the work output, we find that the 
decoupling cost terms 
account for the majority of the reduction for strong coupling. To illustrate this point we have separated these contributions from the remaining work output 
in Fig.~\ref{WorkOutad}. 
For instantaneous decoupling we see that 
even neglecting the cost of switching off the reservoir interactions (dotted curve), the net work extracted along the isentropic strokes is slightly lower than in the weak coupling case (dashed curve). However, the size of the cost term (dashed line) dwarfs this effect, 
to an extent that emphasises just how severe a simplification is made by neglecting 
interaction effects in the weak coupling treatment. We see that the cost can be mitigated to a certain extent by the adiabatic decoupling procedure (solid line). 
As the work extracted neglecting the decoupling cost only recovers 
to the weak coupling limit in this case, the total work output is still lower and so the cost cannot be fully overcome.

\begin{figure}[t]
\includegraphics[width=0.98\textwidth]{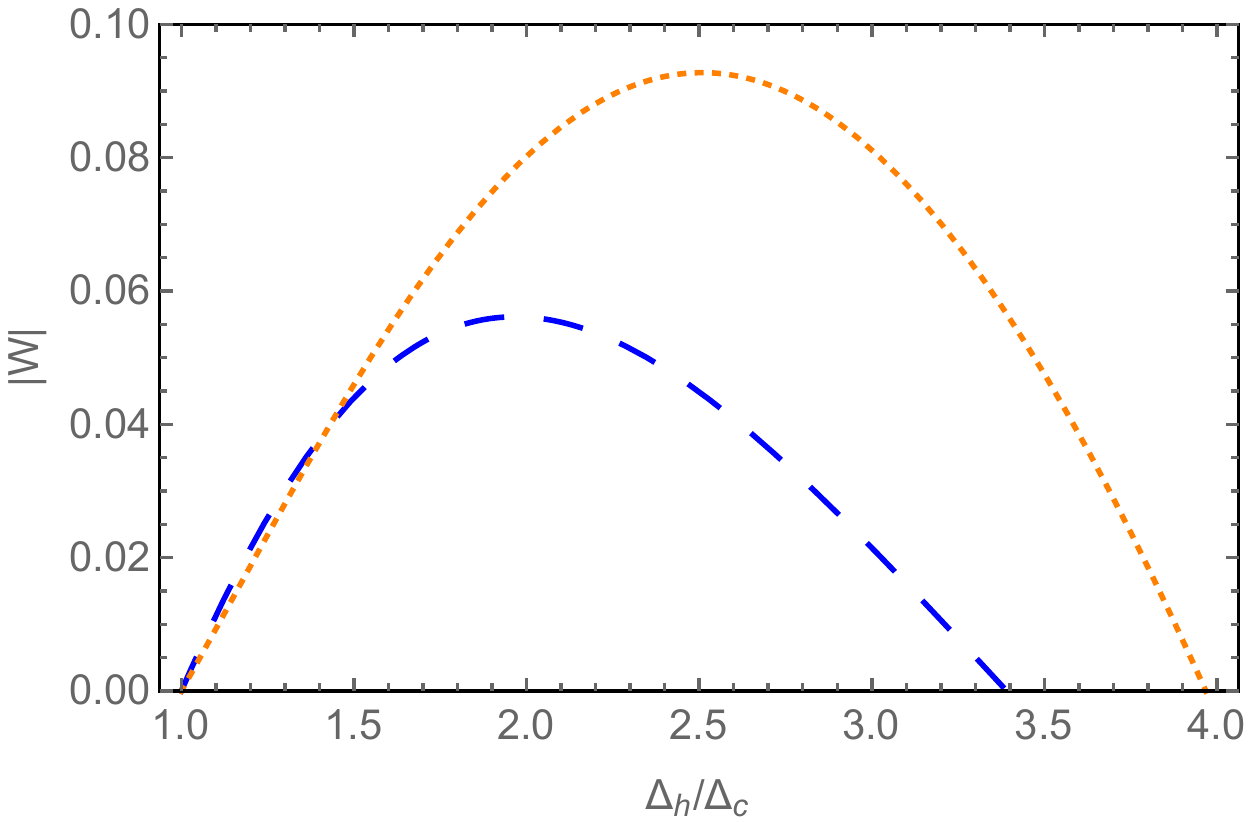}
\caption{{\bf Adiabatic limit.} Work output of the Otto cycle 
 plotted as a function of the TLS tunneling at point $A$, $\Delta_h$, ignoring the decoupling cost for strong coupling (instantaneous decoupling case). \textit{Blue dashed curve:} weak coupling;  \textit{Orange dotted curve:} strong coupling. Parameters (in units of $\Delta_c$): $\epsilon_h=\epsilon_{c}=1$, $\beta_{h}=1$, $\beta_{c}=2.5$, $\omega_{c}=2$, $\alpha=1$, and $n=30$ states are taken in the RC calculations.
\label{WorkOutdeltaad}}
\end{figure}

We show a parametric plot of the work output against the efficiency of the engine for both strong and weak coupling in Fig.~\ref{efficiencyfigad}. The parameter which is varied along these curves is the bias, $\epsilon_h$, at point $A$. 
In the weak coupling regime, the efficiency increases monotonically with $\epsilon_h$ and saturates at the Carnot limit, $\eta_C = 1 - {T_c}/{T_h}$, at the point where the work output vanishes, $\mu_c/\mu_h=T_c/T_h$. The efficiency at maximum work output occurs prior to the Carnot bound being attained. The efficiency of the engine is inferior at strong coupling: work output is reduced in this regime and the energy absorbed from the hot reservoir is not reduced sufficiently to prevent a reduction in efficiency. We observe a qualitatively different behaviour compared with the weak coupling limit: the efficiency is maximised below the Carnot limit before 
turning over and falling to zero as the work output vanishes. This creates the loop structure of both the instantaneous decoupling and adiabatic decoupling curves. This structure is reminiscent of studies of heat engines containing some degree of internal frictional loss, for example in Ref. \cite{1367-2630-17-7-075007}. 
The adiabatic decoupling protocol yields an improved efficiency, with some mitigation of the decoupling costs, and a loop which lies beyond the instantaneous decoupling case.

%We combine the work and energy absorbed to plot the efficiency of the engine for both strong and weak coupling in Fig.~\ref{efficiencyfigad}. 
%In the weak coupling regime, the efficiency increases monotonically with $\epsilon_h$ and saturates at the Carnot limit, $\eta_C = 1 - {T_c}/{T_h}$, at the point where the work output vanishes, $\mu_c/\mu_h=T_c/T_h$ (not shown). As expected, the efficiency of the engine is inferior at strong coupling due to the reduction in work output, with the smaller absorption of energy from the hot reservoir insufficient to offset this loss. It also behaves in a qualitatively different manner to the weak coupling limit,  
%reaching a maximum (below the Carnot limit) before 
%turning over 
%and falling to zero as the work output disappears but the energy absorption remains finite. 
%Again, for adiabatic decoupling of the reservoirs, the efficiency is improved due to the reduction of the associated costs. 

It is worth noting that for instantaneous decoupling in certain parameter regimes, the work output in the strong coupling case can beat the weak coupling limit when ignoring the decoupling cost. An example is shown in Fig.~\ref{WorkOutdeltaad}, 
where we vary $\Delta_h$ along the isentropic strokes rather than $\epsilon_h$. However, even in these situations the decoupling costs far outweigh such enhancements, and 
the cycle always displays 
a reduction in work output for the strong coupling calculations.

\begin{figure}[t]
\vspace{0.1cm}
\includegraphics[width=0.99\textwidth]{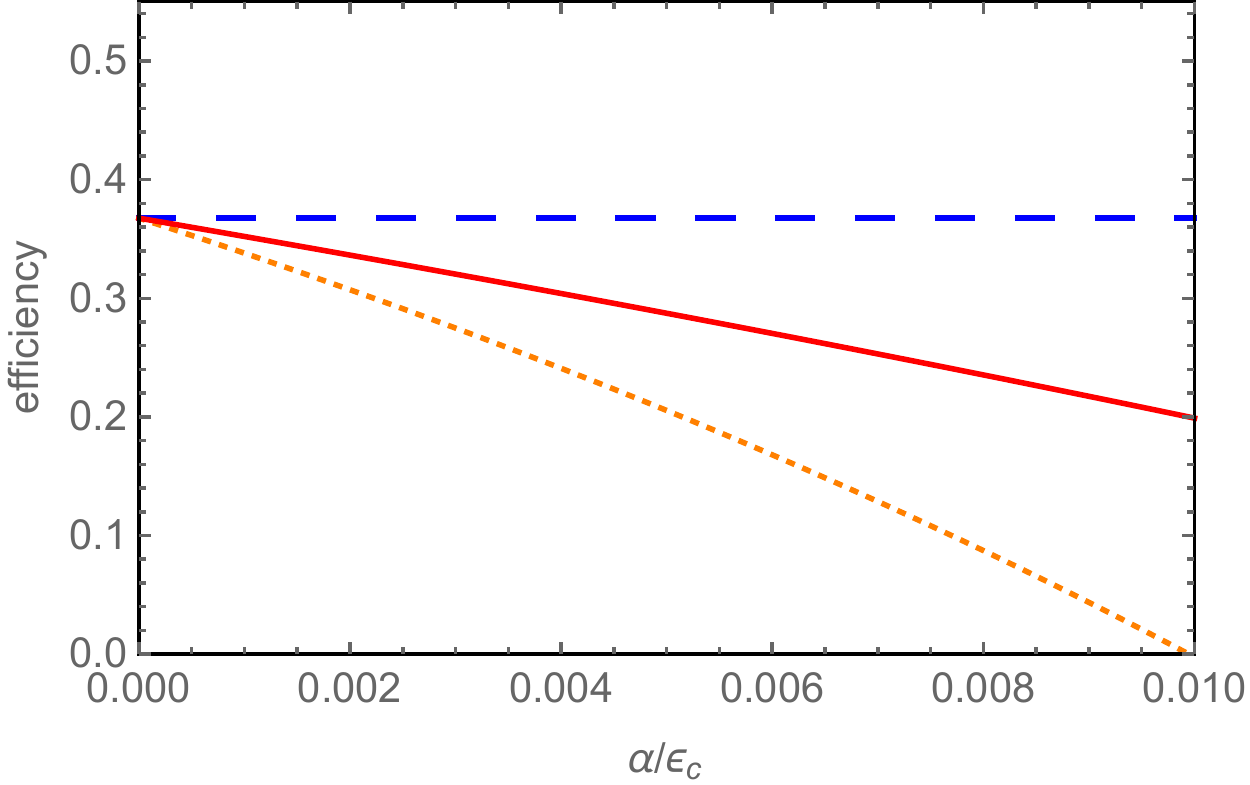}
\caption{{\bf Adiabatic limit.} Efficiency of the Otto engine plotted as a function of coupling strength $\alpha$.  \textit{Blue dashed line:} weak coupling; \textit{Orange dotted curve:} strong coupling with instantaneous decoupling of the reservoirs;  \textit{Red solid curve:} strong coupling with adiabatic decoupling of the reservoirs. Parameters as in Fig.~\ref{Workfigad} with $\epsilon_{h}/\epsilon_c=2$.
\label{efficiencycoupling}}
\end{figure}

Finally, we consider how the engine performance scales with system-reservoir coupling strength $\alpha$ in Fig.~\ref{efficiencycoupling}. 
We observe that as the coupling strength is increased, the engine's performance deteriorates in all but the weak coupling case, since there the magnitude of the interaction term 
is unimportant. Also, as expected, 
the weak coupling efficiency is recovered as $\alpha\rightarrow0$. 
It is clear that the detrimental effect of the decoupling cost is more pronounced at higher couplings, and the adiabatic decoupling procedure becomes increasingly desirable, despite being unable to fully recover the weak coupling efficiency.
In fact, for large enough couplings, it allows the cycle to perform as an engine even when the work output has vanished in the instantaneous decoupling case. 
It is also worth recognising that the sensitivity of the cycle performance to the system-reservoir decoupling procedure (instantaneous or adiabatic) is a feature inherent to the strong coupling regime, i.e.~it is completely absent in the weak coupling treatment. Hence, it could be used (even at fixed $\alpha$) to signify the presence of strong coupling effects in experimental realisations of quantum heat engines.  

\subsection{Sudden isentropic strokes}

We move now to the sudden limit of the Otto cycle, where the isentropes are carried out so quickly that the quantum state has no time to evolve along the stroke. The states at $C$ and $A$ thus read $\chi^{C} = \chi^{B'}$ and $\chi^{A} = \chi^{D'}$, which alters the respective energy expressions at these points. 
Working through the cycle (assuming instantaneous decoupling), we find that in the sudden limit the net work output 
reads
\begin{align}
W  = & \textrm{tr}\left[H_{S}^{B}(\rho_c-\rho_{h})\right]+\textrm{tr}\left[H_{S}^{C}(\rho_h-\rho_{c})\right]\nonumber\\
 &   -\textrm{tr}[H_{I_{h}}\rho_{h}]-\textrm{tr}[H_{I_{c}}\rho_{c}],
 \label{WorkSudden}
\end{align}
while the energy absorbed from the hot reservoir becomes
\begin{align}
Q^{A'B} = & \textrm{tr}\left[H_{S}^{B}(\rho_h-\rho_{c})\right]\nonumber\\
 &+\textrm{tr}\left[H_{R_h}(\rho_{h}-\rho_{R_h})\right]
+\textrm{tr}\left[H_{R_c}(\rho_{R_c}-\rho_{c})\right]\nonumber\\
&+\textrm{tr}[H_{I_{h}}\rho_{h}].
\label{Heat}
\end{align}
Both expressions may be evaluated within the RC formalism, as was the case in the adiabatic treatment.

\begin{figure}[t]
\includegraphics[width=0.96\textwidth]{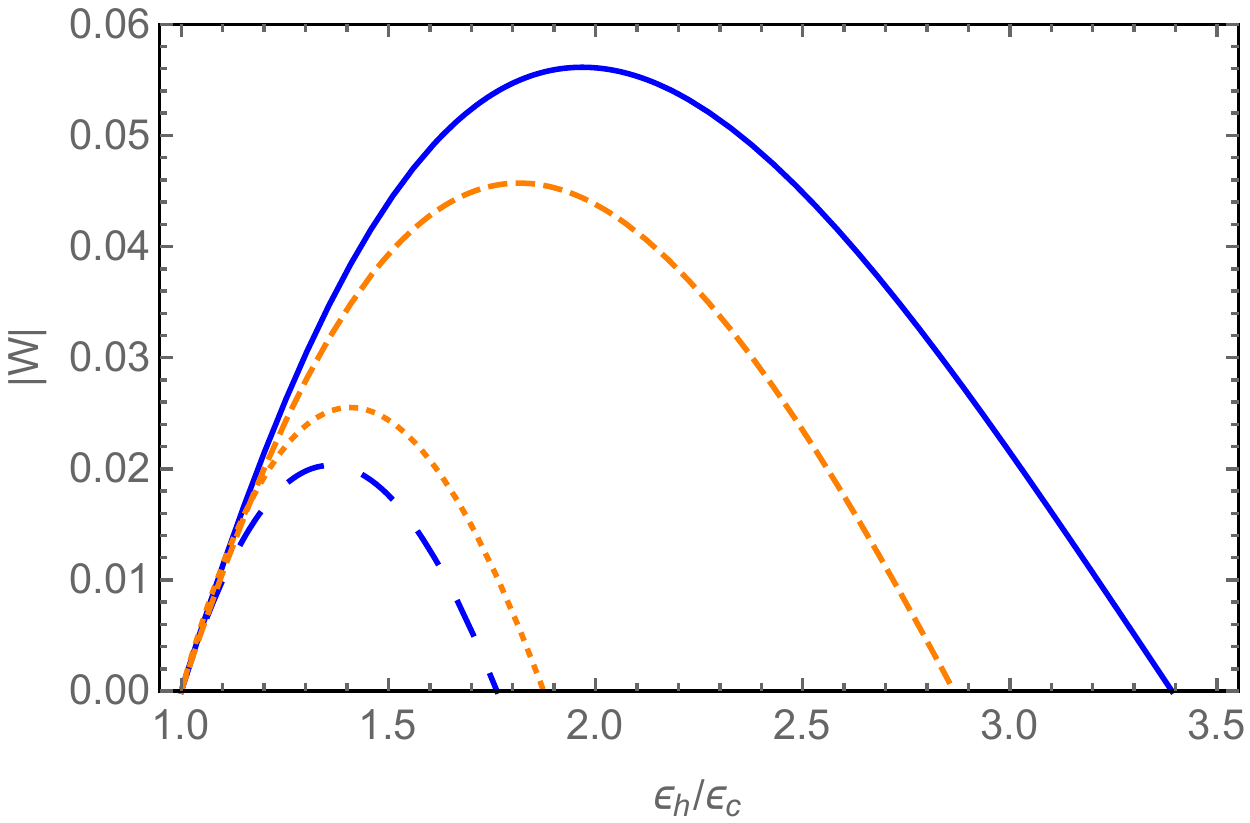}
\caption{{\bf Comparing the adiabatic and sudden limits.} Work output of the Otto cycle 
plotted as a function of the TLS bias at point $A$, $\epsilon_h$, ignoring the decoupling cost for strong coupling. \textit{Blue solid curve:} weak coupling adiabatic; \textit{Orange (small) dashed curve:} strong coupling adiabatic with instantaneous decoupling; \textit{Blue (large) dashed curve:} weak coupling sudden; \textit{Orange dotted curve:} strong coupling sudden.
Parameters as in Fig.~\ref{Workfigad} though with $\alpha/\epsilon_c=0.1$.
\label{WorkOutsud}}
\end{figure}

\begin{figure}[t]
\includegraphics[width=0.99\textwidth]{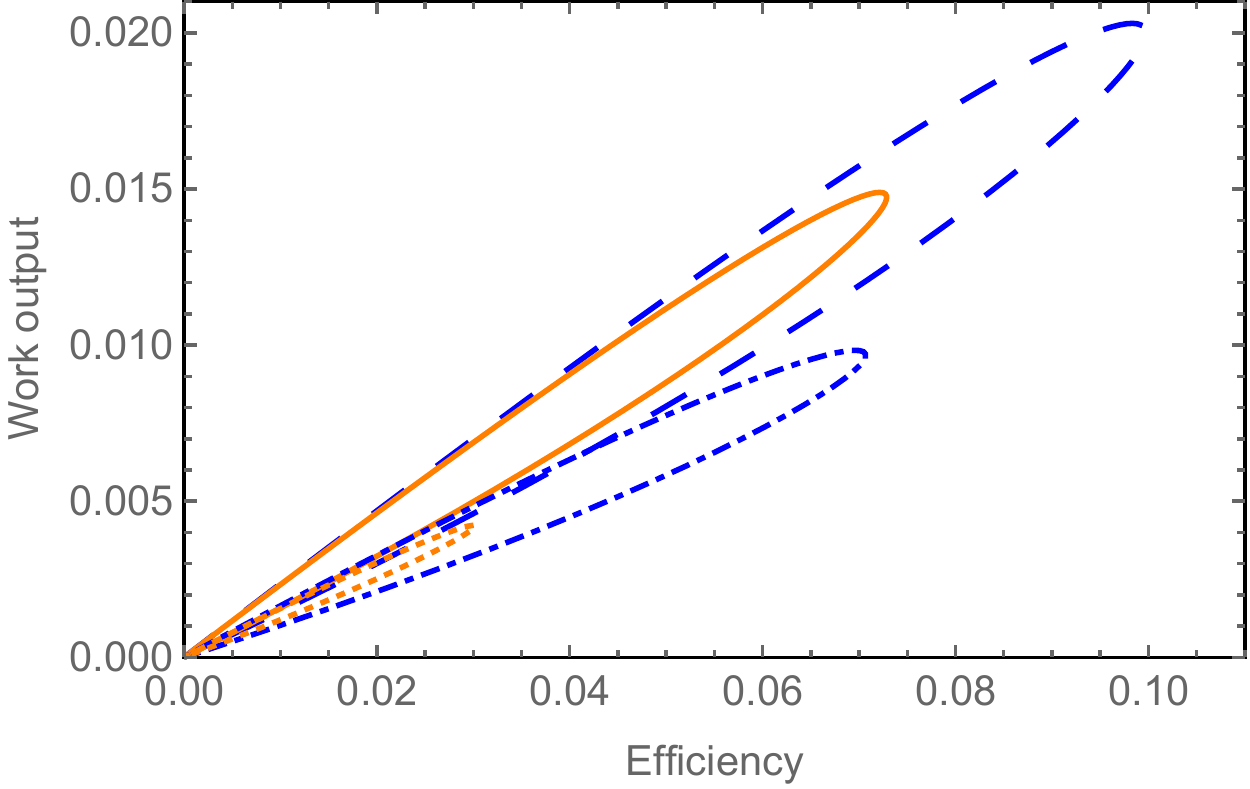}
\caption{{\bf Sudden limit.} Parametric plots of work output against efficiency in the sudden regime for a quantum Otto cycle plotted by varying the TLS bias at point $A$, $\epsilon_h$, for two different temperatures of the cold reservoir. {\it Blue dashed curve}: weak coupling and $\beta_c = 2.5$; {\it Orange solid curve}: strong coupling and $\beta_c = 2.5$; {\it Blue dot-dashed curve}: weak coupling $\beta_c = 1.75$; {\it Orange dotted curve}: strong coupling $\beta_c = 1.75$. Parameters (in units of $\epsilon_c$): $\Delta_h = \Delta_c=1$, $\beta_{h}=1$, $\omega_{c}=2$, $\alpha=0.001$, and $n=30$ states are taken in the RC calculations. 
\label{Workfigsud}}
\end{figure}

It is well known that operating the cycle in the sudden limit introduces a process known as quantum friction \cite{1367-2630-8-5-083, arXiv:1602.06164}, which impacts negatively on the performance of the engine. 
This is apparent, for instance, in Fig.~\ref{WorkOutsud}, where we note a large reduction in the weak coupling work output as compared to the adiabatic limit. Interestingly, we see that in this example the effect of quantum friction is less pronounced in the strong coupling regime, and indeed 
if it were not for the cost of decoupling (which outweighs the benefit), the strong coupling engine would outperform its weakly coupled counterpart. The impacts of the decoupling cost on the work output and energy conversion efficiency are shown in Fig.~\ref{Workfigsud}. In this figure, we see both the effect of quantum friction leading to a loop-like structure qualitatively similar to those seen in Fig. \ref{efficiencyfigad}, and the effect of finite coupling reducing work output and efficiency such that the loops become smaller. We find that the scaling of engine performance with coupling strength is qualitatively similar to the adiabatic isentrope limit in Fig.~\ref{efficiencycoupling}.

\section{Discussion}

We have studied a quantum heat engine in the regime of strong coupling between the system and reservoirs, employing the RC formalism to account for the resulting generation of system-reservoir correlations.  
Considering the quantum Otto cycle, we have shown that the work cost incurred in decoupling the system and reservoirs impacts negatively on the engine, 
and have 
established that a variation of the cycle can help to improve its performance at strong coupling. 

%\begin{figure}[t]
%\includegraphics[width=0.99\textwidth]{sudden_work.pdf}
%\includegraphics[width=0.99\textwidth]{sudden_efficiency.pdf}
%\caption{{\bf Sudden limit.} Work output (top) and efficiency (bottom) of a quantum Otto engine plotted as a function of the TLS bias at point $A$, $\epsilon_h$. \textit{Blue dashed curve:} weak coupling,  \textit{Orange dotted curve:} strong coupling. 
%Parameters as in Fig.~\ref{Workfigad} though with $\alpha/\epsilon_c=0.001$.
%\label{Workfigsud}}
%\end{figure}

At the heart of our treatment lie the TLS-reservoir interactions, which we have modelled 
as being of spin-boson form. 
As such, we had to specify a certain spectral density, 
here chosen to be Ohmic. 
Extending the RC formalism of Refs.~\cite{PhysRevA.90.032114, Iles-Smith2015} to other forms of spectral density is the subject of ongoing work, and should yield further understanding on 
the generality of our results. 
There are also alternative theoretical techniques, for example the polaron transformation~\cite{doi:10.1021/acs.jpclett.5b01404,0953-8984-28-10-103002}, that could be used to perform a similar and complementary analysis. 

Several experimental realisations of nanoscale heat engines have been proposed \cite{Rosnagel2015,PhysRevLett.109.203006,PhysRevB.87.075312,0957-4484-26-3-032001,PhysRevB.91.195406,PhysRevB.93.041418}. Although the protocols presented here are idealised in that they involve complete decoupling of the system from the reservoirs at various points during the engine cycle, it is certainly possible to control very precisely parameters such as the bias and tunnelling, for example, in experimental realisations of double well quantum dots in semiconductor nanowires \cite{Liu285,PhysRevLett.113.036801}. An experimental realisation of a single-atom heat engine has recently been achieved \cite{Rosnagel2015}, indicating that tuning the interaction strength as required for protocols such as those outlined here is possible. The theoretical description of nanoscale quantum engines which do not require decoupling from the reservoirs is given by continuously coupled models, recently reviewed in Ref.~\cite{doi:10.1146/annurev-physchem-040513-103724}. The access afforded by the RC formalism to hitherto unexplored regimes are compelling reasons to apply the techniques discussed here to such models, as in Ref.~\cite{arXiv:1602.01340}. 

For the purposes of this work, we have considered an idealised 
version of the Otto cycle where full equilibration occurs along the isochores and the isentropic strokes are carried out either in the adiabatic or sudden limit. Extensions to finite time versions of the cycle could also be studied, in which case work output and efficiency become functions of the cycle time, and a power output may be defined. 
Considering such a cycle within the RC formalism would enable us to 
evaluate finite power situations at strong coupling, where full equilibration 
does not occur 
along the isochoric strokes. The question of whether strong system-reservoir couplings are also detrimental for the performance of finite time cycles is an important and interesting one, which merits further investigation.

Let us finally comment that the RC formalism has allowed us to consider explicitly the work costs involved in coupling and decoupling 
the system and the reservoirs around a heat engine cycle. These costs are typically ignored in the weak coupling limit where it is assumed that the interaction term in the Hamiltonian 
is small enough that the reservoirs remain in thermal equilibrium, and so 
linear interaction terms may be considered free. This assumption is 
frequently valid in classical thermodynamic systems, 
though at the quantum scale this need not necessarily be so. It may be argued that in considering such costs, we have had to 
specify a particular microscopic model of a heat engine and that, as a result, the emerging thermodynamics becomes model dependent.  As such, one loses the 
appealing generality at the heart of the success of classical thermodynamics. We would argue in response that at strong coupling, whether on the quantum or classical scale, given that the Hamiltonian interaction terms become appreciable, a model specific description of thermodynamics may be inevitable. In which case, a formalism such as the RC method that allows for a physically intuitive treatment of these interactions holds great promise. 

\textit{Acknowledgments---}The authors wish to thank Philipp Strasberg and Neill Lambert for interesting discussions, comments, and for sharing their related manuscript on nonequilibrium thermodynamics in the strong coupling regime prior to publication. We also thank Obinna Abah, Zach Blunden-Codd, and Adam Stokes for discussions. DN is supported by the EPSRC. FM is supported by ERCG grant ODYCQUENT. AN is supported by The University of Manchester and the EPSRC. 

\appendix*
\section{Adiabatic reservoir decoupling}\label{adiabaticdecoupling}
In this Appendix we consider the possibility of mitigating some of the work cost associated with decoupling from the hot and cold reservoirs at points $B'$ and $D'$, respectively. 
If, at the end of the hot isochore, we consider decoupling  
the reservoir from the system 
adiabatically, 
the state at $B'$ then differs from that at $B$ 
and reads
\begin{equation}
\chi^{B'}=\rho_{S_h} \otimes \rho_{th_h} \otimes \rho_{th_c},
\end{equation}
see Eqs.~(\ref{thermalc}) - (\ref{thermalh}). 
Computing the cost of decoupling now yields
\begin{align}
\langle H\rangle^{B'}-\langle H\rangle^{B} = &\textrm{tr}\left[H_{S}^{B}(\rho_{S_h}-\rho_h)\right]+\textrm{tr}\left[H_{R_h}(\rho_{th_h}-\rho_h)\right]\nonumber\\
 & -\textrm{tr}\left[H_{I_{h}}\rho_{h}\right].
 \label{intenergdif}
\end{align}
This cost remains zero in the standard weak coupling analysis. For strong coupling, however, 
it will 
involve both a work contribution 
and heat dissipation into the hot reservoir. We thus compute the change in free energy and equate this to the work associated with decoupling: 
\begin{align} \label{workcost}
F_{B'}-F_{B}=&\langle H\rangle^{B'}-\frac{1}{\beta_h}S(\chi^{B'})-\left[\langle H\rangle^{B}-\frac{1}{\beta_h}S(\chi^{B})\right]\nonumber\\
=& \frac{1}{\beta_{h}}\ln\textrm{tr}\left[\exp\left(-\beta_{h}\left(H_{S}^{B}+H_{R_h}+H_{I_{h}}\right)\right)\right] \nonumber\\
 & -\frac{1}{\beta_{h}}\ln Z_{S_h}-\frac{1}{\beta_{h}}\ln Z_{R_h},
\end{align}
where $S(\rho)=-\textrm{tr}\left[\rho\ln\rho\right]$ is the von Neumann 
entropy, 
$Z_{S_h}=\textrm{tr}\left[\exp\left(-\beta_{h}H_{S}^{B}\right)\right]$, and $Z_{R_h}=\textrm{tr}\left[\exp\left(-\beta_{h}H_{R_h}\right)\right]$. 
The heat dissipated into the bath during decoupling is then 
\begin{eqnarray}
Q_{BB'} & = & \langle H\rangle^{B'}-\langle H\rangle^{B}-\left[F_{B'}-F_{B}\right], 
 \end{eqnarray}
and both the heat and work are calculated within the RC approach. 

We also need to consider decoupling from the cold reservoir at point $D$. If the interaction is switched off adiabatically then the state at $D'$ is given by
\begin{equation}
\chi^{D'}=\rho_{S_c} \otimes \rho_{th_h} \otimes \rho_{th_c},
\end{equation}
where 
\begin{align}\label{rhoSc}
\rho_{S_c} =  \frac{\exp\left(-\beta_{c}H_{S}^C\right)}{\textrm{tr}\left[\exp\left(-\beta_{c}H_{S}^C\right)\right]}.
\end{align}
The cost of decoupling may be evaluated in an analogous fashion to the hot reservoir case to yield
\begin{align}
 \langle H\rangle^{D'}-\langle H\rangle^{D} =&\textrm{tr}\left[H_{S}^{C}(\rho_{S_c}-\rho_{c})\right]+\textrm{tr}\left[H_{R_c} (\rho_{th_c} -\rho_{c})\right]\nonumber\\
& -\textrm{tr}\left[H_{I_{c}}\rho_{c}\right],
 \end{align}
which may be partitioned into a work contribution and energy dissipated into the cold reservoir, just as was done for the hot reservoir.

For adiabatic decoupling, the expression for the net work output around a complete cycle then evaluates to
\begin{eqnarray}
W & = &\left(\frac{\mu_{c}}{\mu_{h}}-1\right)\textrm{tr}\left[H_{S}^{B}\rho_{S_h}\right]+\left(\frac{\mu_{h}}{\mu_{c}}-1\right)\textrm{tr}\left[H_{S}^{C}\rho_{S_c}\right]\nonumber\\
&&+\frac{1}{\beta_{h}}\ln\textrm{tr}\left[\exp\left(-\beta_{h}\left(H_{S}^{B}+H_{R_h}+H_{I_{h}}\right)\right)\right]\nonumber\\
&&+\frac{1}{\beta_{c}}\ln\textrm{tr}\left[\exp\left(-\beta_{c}\left(H_{S}^{C}+H_{R_c}+H_{I_{c}}\right)\right)\right]\nonumber\\
&&-\frac{1}{\beta_{h}}\ln Z_{S_h}-\frac{1}{\beta_{h}}\ln Z_{R_h}\nonumber\\
&&-\frac{1}{\beta_{c}}\ln Z_{S_c}-\frac{1}{\beta_{c}}\ln Z_{R_c}.
\label{Workadec}
\end{eqnarray}
To calculate the energy dissipated into the hot reservoir we need to consider both the contribution from the hot isochore and the subsequent decoupling, $Q = Q_{A'B} +Q_{BB'}$. We then find
\begin{align}
Q  = & -\frac{\mu_{h}}{\mu_{c}}\textrm{tr}\left[H_{S}^{A}\rho_{S_c}\right]-\textrm{tr}\left[H_{S}^{B}\rho_{S_h}\right]\nonumber\\
& +\frac{1}{\beta_{h}}\ln Z_{S_h}+\frac{1}{\beta_{h}}\ln Z_{R_h}\nonumber\\
& -\ln\textrm{tr}\left[\exp\left(-\beta_{h}\left(H_{S}^{B}+H_{R_h}+H_{I_{h}}\right)\right)\right].
\label{Heatadec}
\end{align}

%\bibliography{ottoRCbibliography}

\begin{thebibliography}{61}%
\makeatletter
\providecommand \@ifxundefined [1]{%
 \@ifx{#1\undefined}
}%
\providecommand \@ifnum [1]{%
 \ifnum #1\expandafter \@firstoftwo
 \else \expandafter \@secondoftwo
 \fi
}%
\providecommand \@ifx [1]{%
 \ifx #1\expandafter \@firstoftwo
 \else \expandafter \@secondoftwo
 \fi
}%
\providecommand \natexlab [1]{#1}%
\providecommand \enquote  [1]{``#1''}%
\providecommand \bibnamefont  [1]{#1}%
\providecommand \bibfnamefont [1]{#1}%
\providecommand \citenamefont [1]{#1}%
\providecommand \href@noop [0]{\@secondoftwo}%
\providecommand \href [0]{\begingroup \@sanitize@url \@href}%
\providecommand \@href[1]{\@@startlink{#1}\@@href}%
\providecommand \@@href[1]{\endgroup#1\@@endlink}%
\providecommand \@sanitize@url [0]{\catcode `\\12\catcode `\$12\catcode
  `\&12\catcode `\#12\catcode `\^12\catcode `\_12\catcode `\%12\relax}%
\providecommand \@@startlink[1]{}%
\providecommand \@@endlink[0]{}%
\providecommand \url  [0]{\begingroup\@sanitize@url \@url }%
\providecommand \@url [1]{\endgroup\@href {#1}{\urlprefix }}%
\providecommand \urlprefix  [0]{URL }%
\providecommand \Eprint [0]{\href }%
\providecommand \doibase [0]{http://dx.doi.org/}%
\providecommand \selectlanguage [0]{\@gobble}%
\providecommand \bibinfo  [0]{\@secondoftwo}%
\providecommand \bibfield  [0]{\@secondoftwo}%
\providecommand \translation [1]{[#1]}%
\providecommand \BibitemOpen [0]{}%
\providecommand \bibitemStop [0]{}%
\providecommand \bibitemNoStop [0]{.\EOS\space}%
\providecommand \EOS [0]{\spacefactor3000\relax}%
\providecommand \BibitemShut  [1]{\csname bibitem#1\endcsname}%
\let\auto@bib@innerbib\@empty
%</preamble>
\bibitem [{\citenamefont {Carnot}(1824)}]{Carnot:1824aa}%
  \BibitemOpen
  \bibfield  {author} {\bibinfo {author} {\bibfnamefont {S.}~\bibnamefont
  {Carnot}},\ }\href@noop {} {\emph {\bibinfo {title} {R{\'e}flexions sur la
  puissance motrice du feu: et sur les machines propres {\`a} d{\'e}velopper
  cette puissance}}}\ (\bibinfo  {publisher} {Chez Bachelier, libraire, quai
  des Augustins, no. 55},\ \bibinfo {address} {A Paris},\ \bibinfo {year}
  {1824})\BibitemShut {NoStop}%
\bibitem [{\citenamefont {Blundell}\ and\ \citenamefont
  {Blundell}(2010)}]{Blundell:2010aa}%
  \BibitemOpen
  \bibfield  {author} {\bibinfo {author} {\bibfnamefont {S.}~\bibnamefont
  {Blundell}}\ and\ \bibinfo {author} {\bibfnamefont {K.~M.}\ \bibnamefont
  {Blundell}},\ }\href@noop {} {\emph {\bibinfo {title} {Concepts in thermal
  physics}}},\ \bibinfo {edition} {2nd}\ ed.\ (\bibinfo  {publisher} {Oxford
  University Press},\ \bibinfo {address} {Oxford},\ \bibinfo {year}
  {2010})\BibitemShut {NoStop}%
\bibitem [{\citenamefont {Kosloff}\ and\ \citenamefont
  {Levy}(2014)}]{doi:10.1146/annurev-physchem-040513-103724}%
  \BibitemOpen
  \bibfield  {author} {\bibinfo {author} {\bibfnamefont {R.}~\bibnamefont
  {Kosloff}}\ and\ \bibinfo {author} {\bibfnamefont {A.}~\bibnamefont {Levy}},\
  }\href {\doibase 10.1146/annurev-physchem-040513-103724} {\bibfield
  {journal} {\bibinfo  {journal} {Annual Review of Physical Chemistry}\
  }\textbf {\bibinfo {volume} {65}},\ \bibinfo {pages} {365} (\bibinfo {year}
  {2014})}\BibitemShut {NoStop}%
\bibitem [{\citenamefont {Gelbwaser-Klimovsky}\ \emph
  {et~al.}(2015{\natexlab{a}})\citenamefont {Gelbwaser-Klimovsky},
  \citenamefont {Niedenzu},\ and\ \citenamefont
  {Kurizki}}]{Gelbwaser-Klimovsky2015}%
  \BibitemOpen
  \bibfield  {author} {\bibinfo {author} {\bibfnamefont {D.}~\bibnamefont
  {Gelbwaser-Klimovsky}}, \bibinfo {author} {\bibfnamefont {W.}~\bibnamefont
  {Niedenzu}}, \ and\ \bibinfo {author} {\bibfnamefont {G.}~\bibnamefont
  {Kurizki}},\ }\href {\doibase http://dx.doi.org/10.1016/bs.aamop.2015.07.002}
  {\bibfield  {journal} {\bibinfo  {journal} {Advances In Atomic, Molecular,
  and Optical Physics}\ }\textbf {\bibinfo {volume} {64}},\ \bibinfo {pages}
  {329 } (\bibinfo {year} {2015}{\natexlab{a}})}\BibitemShut {NoStop}%
\bibitem [{\citenamefont {Vinjanampathy}\ and\ \citenamefont
  {Anders}(2016)}]{arXiv:1508.06099}%
  \BibitemOpen
  \bibfield  {author} {\bibinfo {author} {\bibfnamefont {S.}~\bibnamefont
  {Vinjanampathy}}\ and\ \bibinfo {author} {\bibfnamefont {J.}~\bibnamefont
  {Anders}},\ }\href@noop {} {\bibfield  {journal} {\bibinfo  {journal}
  {Contemporary Physics}\ }\textbf {\bibinfo {volume} {57}},\ \bibinfo {pages}
  {1} (\bibinfo {year} {2016})}\BibitemShut {NoStop}%
\bibitem [{\citenamefont {Kosloff}(2013)}]{e15062100}%
  \BibitemOpen
  \bibfield  {author} {\bibinfo {author} {\bibfnamefont {R.}~\bibnamefont
  {Kosloff}},\ }\href {\doibase 10.3390/e15062100} {\bibfield  {journal}
  {\bibinfo  {journal} {Entropy}\ }\textbf {\bibinfo {volume} {15}},\ \bibinfo
  {pages} {2100} (\bibinfo {year} {2013})}\BibitemShut {NoStop}%
\bibitem [{\citenamefont {Skrzypczyk}\ \emph {et~al.}(2014)\citenamefont
  {Skrzypczyk}, \citenamefont {Short},\ and\ \citenamefont
  {Popescu}}]{Skrzypczyk:2014aa}%
  \BibitemOpen
  \bibfield  {author} {\bibinfo {author} {\bibfnamefont {P.}~\bibnamefont
  {Skrzypczyk}}, \bibinfo {author} {\bibfnamefont {A.~J.}\ \bibnamefont
  {Short}}, \ and\ \bibinfo {author} {\bibfnamefont {S.}~\bibnamefont
  {Popescu}},\ }\href {http://dx.doi.org/10.1038/ncomms5185} {\bibfield
  {journal} {\bibinfo  {journal} {Nat Commun}\ }\textbf {\bibinfo {volume} {5}}
  (\bibinfo {year} {2014})}\BibitemShut {NoStop}%
\bibitem [{\citenamefont {Scovil}\ and\ \citenamefont
  {Schulz-DuBois}(1959)}]{PhysRevLett.2.262}%
  \BibitemOpen
  \bibfield  {author} {\bibinfo {author} {\bibfnamefont {H.~E.~D.}\
  \bibnamefont {Scovil}}\ and\ \bibinfo {author} {\bibfnamefont {E.~O.}\
  \bibnamefont {Schulz-DuBois}},\ }\href {\doibase 10.1103/PhysRevLett.2.262}
  {\bibfield  {journal} {\bibinfo  {journal} {Phys. Rev. Lett.}\ }\textbf
  {\bibinfo {volume} {2}},\ \bibinfo {pages} {262} (\bibinfo {year}
  {1959})}\BibitemShut {NoStop}%
\bibitem [{\citenamefont {Quan}\ \emph {et~al.}(2007)\citenamefont {Quan},
  \citenamefont {Liu}, \citenamefont {Sun},\ and\ \citenamefont
  {Nori}}]{PhysRevE.76.031105}%
  \BibitemOpen
  \bibfield  {author} {\bibinfo {author} {\bibfnamefont {H.~T.}\ \bibnamefont
  {Quan}}, \bibinfo {author} {\bibfnamefont {Y.-X.}\ \bibnamefont {Liu}},
  \bibinfo {author} {\bibfnamefont {C.~P.}\ \bibnamefont {Sun}}, \ and\
  \bibinfo {author} {\bibfnamefont {F.}~\bibnamefont {Nori}},\ }\href {\doibase
  10.1103/PhysRevE.76.031105} {\bibfield  {journal} {\bibinfo  {journal} {Phys.
  Rev. E}\ }\textbf {\bibinfo {volume} {76}},\ \bibinfo {pages} {031105}
  (\bibinfo {year} {2007})}\BibitemShut {NoStop}%
\bibitem [{\citenamefont {Kieu}(2004)}]{PhysRevLett.93.140403}%
  \BibitemOpen
  \bibfield  {author} {\bibinfo {author} {\bibfnamefont {T.~D.}\ \bibnamefont
  {Kieu}},\ }\href {\doibase 10.1103/PhysRevLett.93.140403} {\bibfield
  {journal} {\bibinfo  {journal} {Phys. Rev. Lett.}\ }\textbf {\bibinfo
  {volume} {93}},\ \bibinfo {pages} {140403} (\bibinfo {year}
  {2004})}\BibitemShut {NoStop}%
\bibitem [{\citenamefont {Scully}\ \emph {et~al.}(2003)\citenamefont {Scully},
  \citenamefont {Zubairy}, \citenamefont {Agarwal},\ and\ \citenamefont
  {Walther}}]{Scully862}%
  \BibitemOpen
  \bibfield  {author} {\bibinfo {author} {\bibfnamefont {M.~O.}\ \bibnamefont
  {Scully}}, \bibinfo {author} {\bibfnamefont {M.~S.}\ \bibnamefont {Zubairy}},
  \bibinfo {author} {\bibfnamefont {G.~S.}\ \bibnamefont {Agarwal}}, \ and\
  \bibinfo {author} {\bibfnamefont {H.}~\bibnamefont {Walther}},\ }\href
  {\doibase 10.1126/science.1078955} {\bibfield  {journal} {\bibinfo  {journal}
  {Science}\ }\textbf {\bibinfo {volume} {299}},\ \bibinfo {pages} {862}
  (\bibinfo {year} {2003})}\BibitemShut {NoStop}%
\bibitem [{\citenamefont {Scully}(2002)}]{PhysRevLett.88.050602}%
  \BibitemOpen
  \bibfield  {author} {\bibinfo {author} {\bibfnamefont {M.~O.}\ \bibnamefont
  {Scully}},\ }\href {\doibase 10.1103/PhysRevLett.88.050602} {\bibfield
  {journal} {\bibinfo  {journal} {Phys. Rev. Lett.}\ }\textbf {\bibinfo
  {volume} {88}},\ \bibinfo {pages} {050602} (\bibinfo {year}
  {2002})}\BibitemShut {NoStop}%
\bibitem [{\citenamefont {Ro\ss{}nagel}\ \emph {et~al.}(2014)\citenamefont
  {Ro\ss{}nagel}, \citenamefont {Abah}, \citenamefont {Schmidt-Kaler},
  \citenamefont {Singer},\ and\ \citenamefont {Lutz}}]{PhysRevLett.112.030602}%
  \BibitemOpen
  \bibfield  {author} {\bibinfo {author} {\bibfnamefont {J.}~\bibnamefont
  {Ro\ss{}nagel}}, \bibinfo {author} {\bibfnamefont {O.}~\bibnamefont {Abah}},
  \bibinfo {author} {\bibfnamefont {F.}~\bibnamefont {Schmidt-Kaler}}, \bibinfo
  {author} {\bibfnamefont {K.}~\bibnamefont {Singer}}, \ and\ \bibinfo {author}
  {\bibfnamefont {E.}~\bibnamefont {Lutz}},\ }\href {\doibase
  10.1103/PhysRevLett.112.030602} {\bibfield  {journal} {\bibinfo  {journal}
  {Phys. Rev. Lett.}\ }\textbf {\bibinfo {volume} {112}},\ \bibinfo {pages}
  {030602} (\bibinfo {year} {2014})}\BibitemShut {NoStop}%
\bibitem [{\citenamefont {Dillenschneider}\ and\ \citenamefont
  {Lutz}(2009)}]{0295-5075-88-5-50003}%
  \BibitemOpen
  \bibfield  {author} {\bibinfo {author} {\bibfnamefont {R.}~\bibnamefont
  {Dillenschneider}}\ and\ \bibinfo {author} {\bibfnamefont {E.}~\bibnamefont
  {Lutz}},\ }\href {http://stacks.iop.org/0295-5075/88/i=5/a=50003} {\bibfield
  {journal} {\bibinfo  {journal} {EPL (Europhysics Letters)}\ }\textbf
  {\bibinfo {volume} {88}},\ \bibinfo {pages} {50003} (\bibinfo {year}
  {2009})}\BibitemShut {NoStop}%
\bibitem [{\citenamefont {Abah}\ and\ \citenamefont
  {Lutz}(2014)}]{EPL2014Lutz}%
  \BibitemOpen
  \bibfield  {author} {\bibinfo {author} {\bibfnamefont {O.}~\bibnamefont
  {Abah}}\ and\ \bibinfo {author} {\bibfnamefont {E.}~\bibnamefont {Lutz}},\
  }\href {http://stacks.iop.org/0295-5075/106/i=2/a=20001} {\bibfield
  {journal} {\bibinfo  {journal} {EPL (Europhysics Letters)}\ }\textbf
  {\bibinfo {volume} {106}},\ \bibinfo {pages} {20001} (\bibinfo {year}
  {2014})}\BibitemShut {NoStop}%
\bibitem [{\citenamefont {Gelbwaser-Klimovsky}\ \emph
  {et~al.}(2013)\citenamefont {Gelbwaser-Klimovsky}, \citenamefont {Alicki},\
  and\ \citenamefont {Kurizki}}]{PhysRevE.87.012140}%
  \BibitemOpen
  \bibfield  {author} {\bibinfo {author} {\bibfnamefont {D.}~\bibnamefont
  {Gelbwaser-Klimovsky}}, \bibinfo {author} {\bibfnamefont {R.}~\bibnamefont
  {Alicki}}, \ and\ \bibinfo {author} {\bibfnamefont {G.}~\bibnamefont
  {Kurizki}},\ }\href {\doibase 10.1103/PhysRevE.87.012140} {\bibfield
  {journal} {\bibinfo  {journal} {Phys. Rev. E}\ }\textbf {\bibinfo {volume}
  {87}},\ \bibinfo {pages} {012140} (\bibinfo {year} {2013})}\BibitemShut
  {NoStop}%
\bibitem [{\citenamefont {Rezek}\ and\ \citenamefont
  {Kosloff}(2006)}]{1367-2630-8-5-083}%
  \BibitemOpen
  \bibfield  {author} {\bibinfo {author} {\bibfnamefont {Y.}~\bibnamefont
  {Rezek}}\ and\ \bibinfo {author} {\bibfnamefont {R.}~\bibnamefont
  {Kosloff}},\ }\href {http://stacks.iop.org/1367-2630/8/i=5/a=083} {\bibfield
  {journal} {\bibinfo  {journal} {New Journal of Physics}\ }\textbf {\bibinfo
  {volume} {8}},\ \bibinfo {pages} {83} (\bibinfo {year} {2006})}\BibitemShut
  {NoStop}%
\bibitem [{\citenamefont {Friedenberger}\ and\ \citenamefont
  {Lutz}(2015)}]{Friedenberger2015}%
  \BibitemOpen
  \bibfield  {author} {\bibinfo {author} {\bibfnamefont {A.}~\bibnamefont
  {Friedenberger}}\ and\ \bibinfo {author} {\bibfnamefont {E.}~\bibnamefont
  {Lutz}},\ }\href@noop {} {\bibfield  {journal} {\bibinfo  {journal}
  {arXiv:1508.04128}\ } (\bibinfo {year} {2015})}\BibitemShut {NoStop}%
\bibitem [{\citenamefont {Seifert}(2016)}]{PhysRevLett.116.020601}%
  \BibitemOpen
  \bibfield  {author} {\bibinfo {author} {\bibfnamefont {U.}~\bibnamefont
  {Seifert}},\ }\href {\doibase 10.1103/PhysRevLett.116.020601} {\bibfield
  {journal} {\bibinfo  {journal} {Phys. Rev. Lett.}\ }\textbf {\bibinfo
  {volume} {116}},\ \bibinfo {pages} {020601} (\bibinfo {year}
  {2016})}\BibitemShut {NoStop}%
\bibitem [{\citenamefont {S\l{}owik}\ \emph {et~al.}(2013)\citenamefont
  {S\l{}owik}, \citenamefont {Filter}, \citenamefont {Straubel}, \citenamefont
  {Lederer},\ and\ \citenamefont {Rockstuhl}}]{PhysRevB.88.195414}%
  \BibitemOpen
  \bibfield  {author} {\bibinfo {author} {\bibfnamefont {K.}~\bibnamefont
  {S\l{}owik}}, \bibinfo {author} {\bibfnamefont {R.}~\bibnamefont {Filter}},
  \bibinfo {author} {\bibfnamefont {J.}~\bibnamefont {Straubel}}, \bibinfo
  {author} {\bibfnamefont {F.}~\bibnamefont {Lederer}}, \ and\ \bibinfo
  {author} {\bibfnamefont {C.}~\bibnamefont {Rockstuhl}},\ }\href {\doibase
  10.1103/PhysRevB.88.195414} {\bibfield  {journal} {\bibinfo  {journal} {Phys.
  Rev. B}\ }\textbf {\bibinfo {volume} {88}},\ \bibinfo {pages} {195414}
  (\bibinfo {year} {2013})}\BibitemShut {NoStop}%
\bibitem [{\citenamefont {H\"ummer}\ \emph {et~al.}(2013)\citenamefont
  {H\"ummer}, \citenamefont {Garc\'{\i}a-Vidal}, \citenamefont
  {Mart\'{\i}n-Moreno},\ and\ \citenamefont {Zueco}}]{PhysRevB.87.115419}%
  \BibitemOpen
  \bibfield  {author} {\bibinfo {author} {\bibfnamefont {T.}~\bibnamefont
  {H\"ummer}}, \bibinfo {author} {\bibfnamefont {F.~J.}\ \bibnamefont
  {Garc\'{\i}a-Vidal}}, \bibinfo {author} {\bibfnamefont {L.}~\bibnamefont
  {Mart\'{\i}n-Moreno}}, \ and\ \bibinfo {author} {\bibfnamefont
  {D.}~\bibnamefont {Zueco}},\ }\href {\doibase 10.1103/PhysRevB.87.115419}
  {\bibfield  {journal} {\bibinfo  {journal} {Phys. Rev. B}\ }\textbf {\bibinfo
  {volume} {87}},\ \bibinfo {pages} {115419} (\bibinfo {year}
  {2013})}\BibitemShut {NoStop}%
\bibitem [{\citenamefont {Le~Hur}(2012)}]{PhysRevB.85.140506}%
  \BibitemOpen
  \bibfield  {author} {\bibinfo {author} {\bibfnamefont {K.}~\bibnamefont
  {Le~Hur}},\ }\href {\doibase 10.1103/PhysRevB.85.140506} {\bibfield
  {journal} {\bibinfo  {journal} {Phys. Rev. B}\ }\textbf {\bibinfo {volume}
  {85}},\ \bibinfo {pages} {140506} (\bibinfo {year} {2012})}\BibitemShut
  {NoStop}%
\bibitem [{\citenamefont {Goldstein}\ \emph {et~al.}(2013)\citenamefont
  {Goldstein}, \citenamefont {Devoret}, \citenamefont {Houzet},\ and\
  \citenamefont {Glazman}}]{PhysRevLett.110.017002}%
  \BibitemOpen
  \bibfield  {author} {\bibinfo {author} {\bibfnamefont {M.}~\bibnamefont
  {Goldstein}}, \bibinfo {author} {\bibfnamefont {M.~H.}\ \bibnamefont
  {Devoret}}, \bibinfo {author} {\bibfnamefont {M.}~\bibnamefont {Houzet}}, \
  and\ \bibinfo {author} {\bibfnamefont {L.~I.}\ \bibnamefont {Glazman}},\
  }\href {\doibase 10.1103/PhysRevLett.110.017002} {\bibfield  {journal}
  {\bibinfo  {journal} {Phys. Rev. Lett.}\ }\textbf {\bibinfo {volume} {110}},\
  \bibinfo {pages} {017002} (\bibinfo {year} {2013})}\BibitemShut {NoStop}%
\bibitem [{\citenamefont {Peropadre}\ \emph {et~al.}(2013)\citenamefont
  {Peropadre}, \citenamefont {Zueco}, \citenamefont {Porras},\ and\
  \citenamefont {Garc\'{\i}a-Ripoll}}]{PhysRevLett.111.243602}%
  \BibitemOpen
  \bibfield  {author} {\bibinfo {author} {\bibfnamefont {B.}~\bibnamefont
  {Peropadre}}, \bibinfo {author} {\bibfnamefont {D.}~\bibnamefont {Zueco}},
  \bibinfo {author} {\bibfnamefont {D.}~\bibnamefont {Porras}}, \ and\ \bibinfo
  {author} {\bibfnamefont {J.~J.}\ \bibnamefont {Garc\'{\i}a-Ripoll}},\ }\href
  {\doibase 10.1103/PhysRevLett.111.243602} {\bibfield  {journal} {\bibinfo
  {journal} {Phys. Rev. Lett.}\ }\textbf {\bibinfo {volume} {111}},\ \bibinfo
  {pages} {243602} (\bibinfo {year} {2013})}\BibitemShut {NoStop}%
\bibitem [{\citenamefont {Nazir}\ and\ \citenamefont
  {McCutcheon}(2016)}]{0953-8984-28-10-103002}%
  \BibitemOpen
  \bibfield  {author} {\bibinfo {author} {\bibfnamefont {A.}~\bibnamefont
  {Nazir}}\ and\ \bibinfo {author} {\bibfnamefont {D.~P.~S.}\ \bibnamefont
  {McCutcheon}},\ }\href {http://stacks.iop.org/0953-8984/28/i=10/a=103002}
  {\bibfield  {journal} {\bibinfo  {journal} {Journal of Physics: Condensed
  Matter}\ }\textbf {\bibinfo {volume} {28}},\ \bibinfo {pages} {103002}
  (\bibinfo {year} {2016})}\BibitemShut {NoStop}%
\bibitem [{\citenamefont {Wei}\ \emph {et~al.}(2014)\citenamefont {Wei},
  \citenamefont {He}, \citenamefont {He}, \citenamefont {Lu}, \citenamefont
  {Pan}, \citenamefont {Schneider}, \citenamefont {Kamp}, \citenamefont
  {H\"ofling}, \citenamefont {McCutcheon},\ and\ \citenamefont
  {Nazir}}]{PhysRevLett.113.097401}%
  \BibitemOpen
  \bibfield  {author} {\bibinfo {author} {\bibfnamefont {Y.-J.}\ \bibnamefont
  {Wei}}, \bibinfo {author} {\bibfnamefont {Y.}~\bibnamefont {He}}, \bibinfo
  {author} {\bibfnamefont {Y.-M.}\ \bibnamefont {He}}, \bibinfo {author}
  {\bibfnamefont {C.-Y.}\ \bibnamefont {Lu}}, \bibinfo {author} {\bibfnamefont
  {J.-W.}\ \bibnamefont {Pan}}, \bibinfo {author} {\bibfnamefont
  {C.}~\bibnamefont {Schneider}}, \bibinfo {author} {\bibfnamefont
  {M.}~\bibnamefont {Kamp}}, \bibinfo {author} {\bibfnamefont {S.}~\bibnamefont
  {H\"ofling}}, \bibinfo {author} {\bibfnamefont {D.~P.~S.}\ \bibnamefont
  {McCutcheon}}, \ and\ \bibinfo {author} {\bibfnamefont {A.}~\bibnamefont
  {Nazir}},\ }\href {\doibase 10.1103/PhysRevLett.113.097401} {\bibfield
  {journal} {\bibinfo  {journal} {Phys. Rev. Lett.}\ }\textbf {\bibinfo
  {volume} {113}},\ \bibinfo {pages} {097401} (\bibinfo {year}
  {2014})}\BibitemShut {NoStop}%
\bibitem [{\citenamefont {Vogl}\ and\ \citenamefont
  {Weitz}(2009)}]{Vogl:2009aa}%
  \BibitemOpen
  \bibfield  {author} {\bibinfo {author} {\bibfnamefont {U.}~\bibnamefont
  {Vogl}}\ and\ \bibinfo {author} {\bibfnamefont {M.}~\bibnamefont {Weitz}},\
  }\href {http://dx.doi.org/10.1038/nature08203} {\bibfield  {journal}
  {\bibinfo  {journal} {Nature}\ }\textbf {\bibinfo {volume} {461}},\ \bibinfo
  {pages} {70} (\bibinfo {year} {2009})}\BibitemShut {NoStop}%
\bibitem [{\citenamefont {Gelbwaser-Klimovsky}\ \emph
  {et~al.}(2015{\natexlab{b}})\citenamefont {Gelbwaser-Klimovsky},
  \citenamefont {Szczygielski}, \citenamefont {Vogl}, \citenamefont {Sa\ss{}},
  \citenamefont {Alicki}, \citenamefont {Kurizki},\ and\ \citenamefont
  {Weitz}}]{PhysRevA.91.023431}%
  \BibitemOpen
  \bibfield  {author} {\bibinfo {author} {\bibfnamefont {D.}~\bibnamefont
  {Gelbwaser-Klimovsky}}, \bibinfo {author} {\bibfnamefont {K.}~\bibnamefont
  {Szczygielski}}, \bibinfo {author} {\bibfnamefont {U.}~\bibnamefont {Vogl}},
  \bibinfo {author} {\bibfnamefont {A.}~\bibnamefont {Sa\ss{}}}, \bibinfo
  {author} {\bibfnamefont {R.}~\bibnamefont {Alicki}}, \bibinfo {author}
  {\bibfnamefont {G.}~\bibnamefont {Kurizki}}, \ and\ \bibinfo {author}
  {\bibfnamefont {M.}~\bibnamefont {Weitz}},\ }\href {\doibase
  10.1103/PhysRevA.91.023431} {\bibfield  {journal} {\bibinfo  {journal} {Phys.
  Rev. A}\ }\textbf {\bibinfo {volume} {91}},\ \bibinfo {pages} {023431}
  (\bibinfo {year} {2015}{\natexlab{b}})}\BibitemShut {NoStop}%
\bibitem [{Note1()}]{Note1}%
  \BibitemOpen
  \bibinfo {note} {We consider any situation in which the system-reservoir
  factorisation assumption fails to define the strong coupling
  regime.}\BibitemShut {Stop}%
\bibitem [{\citenamefont {Ankerhold}\ and\ \citenamefont
  {Pekola}(2014)}]{PhysRevB.90.075421}%
  \BibitemOpen
  \bibfield  {author} {\bibinfo {author} {\bibfnamefont {J.}~\bibnamefont
  {Ankerhold}}\ and\ \bibinfo {author} {\bibfnamefont {J.~P.}\ \bibnamefont
  {Pekola}},\ }\href {\doibase 10.1103/PhysRevB.90.075421} {\bibfield
  {journal} {\bibinfo  {journal} {Phys. Rev. B}\ }\textbf {\bibinfo {volume}
  {90}},\ \bibinfo {pages} {075421} (\bibinfo {year} {2014})}\BibitemShut
  {NoStop}%
\bibitem [{\citenamefont {Carrega}\ \emph {et~al.}(2015)\citenamefont
  {Carrega}, \citenamefont {Solinas}, \citenamefont {Braggio}, \citenamefont
  {Sassetti},\ and\ \citenamefont {Weiss}}]{1367-2630-17-4-045030}%
  \BibitemOpen
  \bibfield  {author} {\bibinfo {author} {\bibfnamefont {M.}~\bibnamefont
  {Carrega}}, \bibinfo {author} {\bibfnamefont {P.}~\bibnamefont {Solinas}},
  \bibinfo {author} {\bibfnamefont {A.}~\bibnamefont {Braggio}}, \bibinfo
  {author} {\bibfnamefont {M.}~\bibnamefont {Sassetti}}, \ and\ \bibinfo
  {author} {\bibfnamefont {U.}~\bibnamefont {Weiss}},\ }\href
  {http://stacks.iop.org/1367-2630/17/i=4/a=045030} {\bibfield  {journal}
  {\bibinfo  {journal} {New Journal of Physics}\ }\textbf {\bibinfo {volume}
  {17}},\ \bibinfo {pages} {045030} (\bibinfo {year} {2015})}\BibitemShut
  {NoStop}%
\bibitem [{\citenamefont {Esposito}\ \emph {et~al.}(2015)\citenamefont
  {Esposito}, \citenamefont {Ochoa},\ and\ \citenamefont
  {Galperin}}]{PhysRevLett.114.080602}%
  \BibitemOpen
  \bibfield  {author} {\bibinfo {author} {\bibfnamefont {M.}~\bibnamefont
  {Esposito}}, \bibinfo {author} {\bibfnamefont {M.~A.}\ \bibnamefont {Ochoa}},
  \ and\ \bibinfo {author} {\bibfnamefont {M.}~\bibnamefont {Galperin}},\
  }\href {\doibase 10.1103/PhysRevLett.114.080602} {\bibfield  {journal}
  {\bibinfo  {journal} {Phys. Rev. Lett.}\ }\textbf {\bibinfo {volume} {114}},\
  \bibinfo {pages} {080602} (\bibinfo {year} {2015})}\BibitemShut {NoStop}%
\bibitem [{\citenamefont {Campisi}\ \emph {et~al.}(2009)\citenamefont
  {Campisi}, \citenamefont {Talkner},\ and\ \citenamefont
  {H\"anggi}}]{PhysRevLett.102.210401}%
  \BibitemOpen
  \bibfield  {author} {\bibinfo {author} {\bibfnamefont {M.}~\bibnamefont
  {Campisi}}, \bibinfo {author} {\bibfnamefont {P.}~\bibnamefont {Talkner}}, \
  and\ \bibinfo {author} {\bibfnamefont {P.}~\bibnamefont {H\"anggi}},\ }\href
  {\doibase 10.1103/PhysRevLett.102.210401} {\bibfield  {journal} {\bibinfo
  {journal} {Phys. Rev. Lett.}\ }\textbf {\bibinfo {volume} {102}},\ \bibinfo
  {pages} {210401} (\bibinfo {year} {2009})}\BibitemShut {NoStop}%
\bibitem [{\citenamefont {Campisi}\ \emph {et~al.}(2011)\citenamefont
  {Campisi}, \citenamefont {H\"anggi},\ and\ \citenamefont
  {Talkner}}]{RevModPhys.83.771}%
  \BibitemOpen
  \bibfield  {author} {\bibinfo {author} {\bibfnamefont {M.}~\bibnamefont
  {Campisi}}, \bibinfo {author} {\bibfnamefont {P.}~\bibnamefont {H\"anggi}}, \
  and\ \bibinfo {author} {\bibfnamefont {P.}~\bibnamefont {Talkner}},\ }\href
  {\doibase 10.1103/RevModPhys.83.771} {\bibfield  {journal} {\bibinfo
  {journal} {Rev. Mod. Phys.}\ }\textbf {\bibinfo {volume} {83}},\ \bibinfo
  {pages} {771} (\bibinfo {year} {2011})}\BibitemShut {NoStop}%
\bibitem [{\citenamefont {Hanggi}\ and\ \citenamefont
  {Talkner}(2015)}]{Hanggi:2015aa}%
  \BibitemOpen
  \bibfield  {author} {\bibinfo {author} {\bibfnamefont {P.}~\bibnamefont
  {Hanggi}}\ and\ \bibinfo {author} {\bibfnamefont {P.}~\bibnamefont
  {Talkner}},\ }\href {http://dx.doi.org/10.1038/nphys3167} {\bibfield
  {journal} {\bibinfo  {journal} {Nat Phys}\ }\textbf {\bibinfo {volume}
  {11}},\ \bibinfo {pages} {108} (\bibinfo {year} {2015})}\BibitemShut
  {NoStop}%
\bibitem [{\citenamefont {H{\"o}rhammer}\ and\ \citenamefont
  {B{\"u}ttner}(2008)}]{Horhammer2008}%
  \BibitemOpen
  \bibfield  {author} {\bibinfo {author} {\bibfnamefont {C.}~\bibnamefont
  {H{\"o}rhammer}}\ and\ \bibinfo {author} {\bibfnamefont {H.}~\bibnamefont
  {B{\"u}ttner}},\ }\href {\doibase 10.1007/s10955-008-9640-x} {\bibfield
  {journal} {\bibinfo  {journal} {Journal of Statistical Physics}\ }\textbf
  {\bibinfo {volume} {133}},\ \bibinfo {pages} {1161} (\bibinfo {year}
  {2008})}\BibitemShut {NoStop}%
\bibitem [{\citenamefont {Esposito}\ \emph {et~al.}(2010)\citenamefont
  {Esposito}, \citenamefont {Lindenberg},\ and\ \citenamefont {den
  Broeck}}]{1367-2630-12-1-013013}%
  \BibitemOpen
  \bibfield  {author} {\bibinfo {author} {\bibfnamefont {M.}~\bibnamefont
  {Esposito}}, \bibinfo {author} {\bibfnamefont {K.}~\bibnamefont
  {Lindenberg}}, \ and\ \bibinfo {author} {\bibfnamefont {C.~V.}\ \bibnamefont
  {den Broeck}},\ }\href {http://stacks.iop.org/1367-2630/12/i=1/a=013013}
  {\bibfield  {journal} {\bibinfo  {journal} {New Journal of Physics}\ }\textbf
  {\bibinfo {volume} {12}},\ \bibinfo {pages} {013013} (\bibinfo {year}
  {2010})}\BibitemShut {NoStop}%
\bibitem [{\citenamefont {Deffner}\ and\ \citenamefont
  {Lutz}(2011)}]{PhysRevLett.107.140404}%
  \BibitemOpen
  \bibfield  {author} {\bibinfo {author} {\bibfnamefont {S.}~\bibnamefont
  {Deffner}}\ and\ \bibinfo {author} {\bibfnamefont {E.}~\bibnamefont {Lutz}},\
  }\href {\doibase 10.1103/PhysRevLett.107.140404} {\bibfield  {journal}
  {\bibinfo  {journal} {Phys. Rev. Lett.}\ }\textbf {\bibinfo {volume} {107}},\
  \bibinfo {pages} {140404} (\bibinfo {year} {2011})}\BibitemShut {NoStop}%
\bibitem [{\citenamefont {Pucci}\ \emph {et~al.}(2013)\citenamefont {Pucci},
  \citenamefont {Esposito},\ and\ \citenamefont
  {Peliti}}]{1742-5468-2013-04-P04005}%
  \BibitemOpen
  \bibfield  {author} {\bibinfo {author} {\bibfnamefont {L.}~\bibnamefont
  {Pucci}}, \bibinfo {author} {\bibfnamefont {M.}~\bibnamefont {Esposito}}, \
  and\ \bibinfo {author} {\bibfnamefont {L.}~\bibnamefont {Peliti}},\ }\href
  {http://stacks.iop.org/1742-5468/2013/i=04/a=P04005} {\bibfield  {journal}
  {\bibinfo  {journal} {Journal of Statistical Mechanics: Theory and
  Experiment}\ }\textbf {\bibinfo {volume} {2013}},\ \bibinfo {pages} {P04005}
  (\bibinfo {year} {2013})}\BibitemShut {NoStop}%
\bibitem [{\citenamefont {Gelbwaser-Klimovsky}\ and\ \citenamefont
  {Aspuru-Guzik}(2015)}]{doi:10.1021/acs.jpclett.5b01404}%
  \BibitemOpen
  \bibfield  {author} {\bibinfo {author} {\bibfnamefont {D.}~\bibnamefont
  {Gelbwaser-Klimovsky}}\ and\ \bibinfo {author} {\bibfnamefont
  {A.}~\bibnamefont {Aspuru-Guzik}},\ }\href {\doibase
  10.1021/acs.jpclett.5b01404} {\bibfield  {journal} {\bibinfo  {journal} {The
  Journal of Physical Chemistry Letters}\ }\textbf {\bibinfo {volume} {6}},\
  \bibinfo {pages} {3477} (\bibinfo {year} {2015})}\BibitemShut {NoStop}%
\bibitem [{\citenamefont {Strasberg}\ \emph {et~al.}(2016)\citenamefont
  {Strasberg}, \citenamefont {Schaller}, \citenamefont {Lambert},\ and\
  \citenamefont {Brandes}}]{arXiv:1602.01340}%
  \BibitemOpen
  \bibfield  {author} {\bibinfo {author} {\bibfnamefont {P.}~\bibnamefont
  {Strasberg}}, \bibinfo {author} {\bibfnamefont {G.}~\bibnamefont {Schaller}},
  \bibinfo {author} {\bibfnamefont {N.}~\bibnamefont {Lambert}}, \ and\
  \bibinfo {author} {\bibfnamefont {T.}~\bibnamefont {Brandes}},\ }\href
  {http://stacks.iop.org/1367-2630/18/i=7/a=073007} {\bibfield  {journal}
  {\bibinfo  {journal} {New Journal of Physics}\ }\textbf {\bibinfo {volume}
  {18}},\ \bibinfo {pages} {073007} (\bibinfo {year} {2016})}\BibitemShut
  {NoStop}%
\bibitem [{\citenamefont {Katz}\ and\ \citenamefont
  {Kosloff}(2016)}]{e18050186}%
  \BibitemOpen
  \bibfield  {author} {\bibinfo {author} {\bibfnamefont {G.}~\bibnamefont
  {Katz}}\ and\ \bibinfo {author} {\bibfnamefont {R.}~\bibnamefont {Kosloff}},\
  }\href {\doibase 10.3390/e18050186} {\bibfield  {journal} {\bibinfo
  {journal} {Entropy}\ }\textbf {\bibinfo {volume} {18}},\ \bibinfo {pages}
  {186} (\bibinfo {year} {2016})}\BibitemShut {NoStop}%
\bibitem [{\citenamefont {Gallego}\ \emph {et~al.}(2014)\citenamefont
  {Gallego}, \citenamefont {Riera},\ and\ \citenamefont
  {Eisert}}]{1367-2630-16-12-125009}%
  \BibitemOpen
  \bibfield  {author} {\bibinfo {author} {\bibfnamefont {R.}~\bibnamefont
  {Gallego}}, \bibinfo {author} {\bibfnamefont {A.}~\bibnamefont {Riera}}, \
  and\ \bibinfo {author} {\bibfnamefont {J.}~\bibnamefont {Eisert}},\ }\href
  {http://stacks.iop.org/1367-2630/16/i=12/a=125009} {\bibfield  {journal}
  {\bibinfo  {journal} {New Journal of Physics}\ }\textbf {\bibinfo {volume}
  {16}},\ \bibinfo {pages} {125009} (\bibinfo {year} {2014})}\BibitemShut
  {NoStop}%
\bibitem [{\citenamefont {Leggett}\ \emph {et~al.}(1987)\citenamefont
  {Leggett}, \citenamefont {Chakravarty}, \citenamefont {Dorsey}, \citenamefont
  {Fisher}, \citenamefont {Garg},\ and\ \citenamefont
  {Zwerger}}]{RevModPhys.59.1}%
  \BibitemOpen
  \bibfield  {author} {\bibinfo {author} {\bibfnamefont {A.~J.}\ \bibnamefont
  {Leggett}}, \bibinfo {author} {\bibfnamefont {S.}~\bibnamefont
  {Chakravarty}}, \bibinfo {author} {\bibfnamefont {A.~T.}\ \bibnamefont
  {Dorsey}}, \bibinfo {author} {\bibfnamefont {M.~P.~A.}\ \bibnamefont
  {Fisher}}, \bibinfo {author} {\bibfnamefont {A.}~\bibnamefont {Garg}}, \ and\
  \bibinfo {author} {\bibfnamefont {W.}~\bibnamefont {Zwerger}},\ }\href
  {\doibase 10.1103/RevModPhys.59.1} {\bibfield  {journal} {\bibinfo  {journal}
  {Rev. Mod. Phys.}\ }\textbf {\bibinfo {volume} {59}},\ \bibinfo {pages} {1}
  (\bibinfo {year} {1987})}\BibitemShut {NoStop}%
\bibitem [{\citenamefont {Weiss}(2012)}]{weissbook}%
  \BibitemOpen
  \bibfield  {author} {\bibinfo {author} {\bibfnamefont {U.}~\bibnamefont
  {Weiss}},\ }\href {https://books.google.co.uk/books?id=qgfuFZxvGKQC} {\emph
  {\bibinfo {title} {Quantum Dissipative Systems}}}\ (\bibinfo  {publisher}
  {World Scientific},\ \bibinfo {year} {2012})\BibitemShut {NoStop}%
\bibitem [{\citenamefont {Nitzan}(2013)}]{nitzanbook}%
  \BibitemOpen
  \bibfield  {author} {\bibinfo {author} {\bibfnamefont {A.}~\bibnamefont
  {Nitzan}},\ }\href {https://books.google.co.uk/books?id=jrmznAEACAAJ} {\emph
  {\bibinfo {title} {Chemical Dynamics in Condensed Phases: Relaxation,
  Transfer, and Reactions in Condensed Molecular Systems}}}\ (\bibinfo
  {publisher} {OUP Oxford},\ \bibinfo {year} {2013})\BibitemShut {NoStop}%
\bibitem [{\citenamefont {Breuer}\ and\ \citenamefont
  {Petruccione}(2002)}]{Breuer:2002aa}%
  \BibitemOpen
  \bibfield  {author} {\bibinfo {author} {\bibfnamefont {H.-P.}\ \bibnamefont
  {Breuer}}\ and\ \bibinfo {author} {\bibfnamefont {F.}~\bibnamefont
  {Petruccione}},\ }\href
  {http://www.loc.gov/catdir/enhancements/fy0613/2002075713-d.html} {\emph
  {\bibinfo {title} {The theory of open quantum systems}}}\ (\bibinfo
  {publisher} {Oxford University Press},\ \bibinfo {address} {Oxford},\
  \bibinfo {year} {2002})\BibitemShut {NoStop}%
\bibitem [{\citenamefont {Schlosshauer}(2007)}]{Schlosshauer:2007aa}%
  \BibitemOpen
  \bibfield  {author} {\bibinfo {author} {\bibfnamefont {M.~A.}\ \bibnamefont
  {Schlosshauer}},\ }\href
  {http://www.loc.gov/catdir/enhancements/fy0814/2007930038-b.html} {\emph
  {\bibinfo {title} {Decoherence and the quantum-to-classical transition}}}\
  (\bibinfo  {publisher} {Springer},\ \bibinfo {address} {Berlin},\ \bibinfo
  {year} {2007})\BibitemShut {NoStop}%
\bibitem [{\citenamefont {Iles-Smith}\ \emph {et~al.}(2014)\citenamefont
  {Iles-Smith}, \citenamefont {Lambert},\ and\ \citenamefont
  {Nazir}}]{PhysRevA.90.032114}%
  \BibitemOpen
  \bibfield  {author} {\bibinfo {author} {\bibfnamefont {J.}~\bibnamefont
  {Iles-Smith}}, \bibinfo {author} {\bibfnamefont {N.}~\bibnamefont {Lambert}},
  \ and\ \bibinfo {author} {\bibfnamefont {A.}~\bibnamefont {Nazir}},\ }\href
  {\doibase 10.1103/PhysRevA.90.032114} {\bibfield  {journal} {\bibinfo
  {journal} {Phys. Rev. A}\ }\textbf {\bibinfo {volume} {90}},\ \bibinfo
  {pages} {032114} (\bibinfo {year} {2014})}\BibitemShut {NoStop}%
\bibitem [{\citenamefont {Liu}\ \emph {et~al.}(2015)\citenamefont {Liu},
  \citenamefont {Stehlik}, \citenamefont {Eichler}, \citenamefont {Gullans},
  \citenamefont {Taylor},\ and\ \citenamefont {Petta}}]{Liu285}%
  \BibitemOpen
  \bibfield  {author} {\bibinfo {author} {\bibfnamefont {Y.-Y.}\ \bibnamefont
  {Liu}}, \bibinfo {author} {\bibfnamefont {J.}~\bibnamefont {Stehlik}},
  \bibinfo {author} {\bibfnamefont {C.}~\bibnamefont {Eichler}}, \bibinfo
  {author} {\bibfnamefont {M.~J.}\ \bibnamefont {Gullans}}, \bibinfo {author}
  {\bibfnamefont {J.~M.}\ \bibnamefont {Taylor}}, \ and\ \bibinfo {author}
  {\bibfnamefont {J.~R.}\ \bibnamefont {Petta}},\ }\href {\doibase
  10.1126/science.aaa2501} {\bibfield  {journal} {\bibinfo  {journal}
  {Science}\ }\textbf {\bibinfo {volume} {347}},\ \bibinfo {pages} {285}
  (\bibinfo {year} {2015})}\BibitemShut {NoStop}%
\bibitem [{\citenamefont {Liu}\ \emph {et~al.}(2014)\citenamefont {Liu},
  \citenamefont {Petersson}, \citenamefont {Stehlik}, \citenamefont {Taylor},\
  and\ \citenamefont {Petta}}]{PhysRevLett.113.036801}%
  \BibitemOpen
  \bibfield  {author} {\bibinfo {author} {\bibfnamefont {Y.-Y.}\ \bibnamefont
  {Liu}}, \bibinfo {author} {\bibfnamefont {K.~D.}\ \bibnamefont {Petersson}},
  \bibinfo {author} {\bibfnamefont {J.}~\bibnamefont {Stehlik}}, \bibinfo
  {author} {\bibfnamefont {J.~M.}\ \bibnamefont {Taylor}}, \ and\ \bibinfo
  {author} {\bibfnamefont {J.~R.}\ \bibnamefont {Petta}},\ }\href {\doibase
  10.1103/PhysRevLett.113.036801} {\bibfield  {journal} {\bibinfo  {journal}
  {Phys. Rev. Lett.}\ }\textbf {\bibinfo {volume} {113}},\ \bibinfo {pages}
  {036801} (\bibinfo {year} {2014})}\BibitemShut {NoStop}%
\bibitem [{\citenamefont {Gilmore}\ \emph {et~al.}(2005)\citenamefont
  {Gilmore}, \ and\ \citenamefont
  {McKenzie}}]{Gilmore2005}%
  \BibitemOpen
  \bibfield  {author} {\bibinfo {author} {\bibfnamefont {J.}~\bibnamefont
  {Gilmore}},
  \ and\ \bibinfo {author} {\bibfnamefont {R.}~\bibnamefont {McKenzie}},\ }\href
  {\doibase } {\bibfield  {journal} {\bibinfo
  {journal} {Journal of Physics: Condensed Matter}\ }\textbf {\bibinfo {volume} {17}},\ \bibinfo
  {pages} {1735} (\bibinfo {year} {2005})}\BibitemShut {NoStop}%  
\bibitem [{\citenamefont {Gullans}\ \emph {et~al.}(2015)\citenamefont
  {Gullans}, \citenamefont {Liu}, \citenamefont {Stehlik}, \citenamefont
  {Petta},\ and\ \citenamefont {Taylor}}]{PhysRevLett.114.196802}%
  \BibitemOpen
  \bibfield  {author} {\bibinfo {author} {\bibfnamefont {M.~J.}\ \bibnamefont
  {Gullans}}, \bibinfo {author} {\bibfnamefont {Y.-Y.}\ \bibnamefont {Liu}},
  \bibinfo {author} {\bibfnamefont {J.}~\bibnamefont {Stehlik}}, \bibinfo
  {author} {\bibfnamefont {J.~R.}\ \bibnamefont {Petta}}, \ and\ \bibinfo
  {author} {\bibfnamefont {J.~M.}\ \bibnamefont {Taylor}},\ }\href {\doibase
  10.1103/PhysRevLett.114.196802} {\bibfield  {journal} {\bibinfo  {journal}
  {Phys. Rev. Lett.}\ }\textbf {\bibinfo {volume} {114}},\ \bibinfo {pages}
  {196802} (\bibinfo {year} {2015})}\BibitemShut {NoStop}%
\bibitem [{\citenamefont {Shevchenko}\ \emph {et~al.}(2010)\citenamefont
  {Shevchenko}, \citenamefont {Ashhab},\ and\ \citenamefont
  {Nori}}]{Shevchenko20101}%
  \BibitemOpen
  \bibfield  {author} {\bibinfo {author} {\bibfnamefont {S.}~\bibnamefont
  {Shevchenko}}, \bibinfo {author} {\bibfnamefont {S.}~\bibnamefont {Ashhab}},
  \ and\ \bibinfo {author} {\bibfnamefont {F.}~\bibnamefont {Nori}},\ }\href
  {\doibase http://dx.doi.org/10.1016/j.physrep.2010.03.002} {\bibfield
  {journal} {\bibinfo  {journal} {Physics Reports}\ }\textbf {\bibinfo {volume}
  {492}},\ \bibinfo {pages} {1 } (\bibinfo {year} {2010})}\BibitemShut
  {NoStop}%
\bibitem [{\citenamefont {Iles-Smith}\ \emph {et~al.}(2016)\citenamefont
  {Iles-Smith}, \citenamefont {Dijkstra}, \citenamefont {Lambert},\ and\
  \citenamefont {Nazir}}]{Iles-Smith2015}%
  \BibitemOpen
  \bibfield  {author} {\bibinfo {author} {\bibfnamefont {J.}~\bibnamefont
  {Iles-Smith}}, \bibinfo {author} {\bibfnamefont {A.~G.}\ \bibnamefont
  {Dijkstra}}, \bibinfo {author} {\bibfnamefont {N.}~\bibnamefont {Lambert}}, \
  and\ \bibinfo {author} {\bibfnamefont {A.}~\bibnamefont {Nazir}},\ }\href
  {http://scitation.aip.org/content/aip/journal/jcp/144/4/10.1063/1.4940218}
  {\bibfield  {journal} {\bibinfo  {journal} {The Journal of Chemical Physics}\
  }\textbf {\bibinfo {volume} {144}},\ \bibinfo {eid} {044110} (\bibinfo {year}
  {2016})}\BibitemShut {NoStop}%
\bibitem [{Note2()}]{Note2}%
  \BibitemOpen
  \bibinfo {note} {The mapped Hamiltonian defined in Eq.~(\ref{RC hamiltonian}) differs from
  that given in  Eq.~$(2)$ of Ref.~\onlinecite{PhysRevA.90.032114} in that it does not include an extra term that is quadratic in the RC creation and annihilation operators, and the couplings $g_k$. As explained in Ref.~\cite{PhysRevA.90.032114}, there
  the term is added to cancel energy shifts that appear in the master equation governing the dynamics of the enlarged
  system. However, as we shall be concerned here with steady state solutions for the enlarged system, which are independent of the quadratic term, it has no bearing in our case, and we work directly with the mapped Hamiltonian as given in Eq.~(\ref{RC hamiltonian}).}\BibitemShut {Stop}%
\bibitem [{\citenamefont {Alecce}\ \emph {et~al.}(2015)\citenamefont
  {Alecce}, \citenamefont {Galve}, \citenamefont {Lo Gullo},
  \citenamefont {Dell'Anna}, \citenamefont {Plastina}, \ and\ \citenamefont {Zambrini}}]{1367-2630-17-7-075007}%
  \BibitemOpen
  \bibfield  {author} {\bibinfo {author} {\bibfnamefont {A.}~\bibnamefont
  {Alecce}}, \bibinfo {author} {\bibfnamefont {F.}\ \bibnamefont
  {Galve}}, \bibinfo {author} {\bibfnamefont {N.}~\bibnamefont {Lo Gullo}}, \bibinfo {author} {\bibfnamefont {L.}~\bibnamefont {Dell'Anna}}, \bibinfo {author} {\bibfnamefont {F.}~\bibnamefont {Plastina}}, \
  and\ \bibinfo {author} {\bibfnamefont {R.}~\bibnamefont {Zambrini}},\ }\href
  {http://iopscience.iop.org/article/10.1088/1367-2630/17/7/075007}
  {\bibfield  {journal} {\bibinfo  {journal} {New Journal of Physics}\
  }\textbf {\bibinfo {volume} {17}},\ \bibinfo {eid} {075007} (\bibinfo {year}
  {2015})}\BibitemShut {NoStop}%  
\bibitem [{\citenamefont {{\c C}akmak}\ \emph {et~al.}(2016)\citenamefont {{\c
  C}akmak}, \citenamefont {F.Altintas},\ and\ \citenamefont
  {M{\"u}stecaplõo{\u{g}}lu}}]{arXiv:1602.06164}%
  \BibitemOpen
  \bibfield  {author} {\bibinfo {author} {\bibfnamefont {S.}~\bibnamefont {{\c
  C}akmak}}, \bibinfo {author} {\bibnamefont {F.Altintas}}, \ and\ \bibinfo
  {author} {\bibfnamefont {\"{O}.~E.}\ \bibnamefont {M{\"u}stecaplio{\u{g}}lu}},\ } {\bibfield
  {journal} {\bibinfo  {journal} {Physica Scripta}\ }\textbf {\bibinfo {volume}
  {91}},\ \bibinfo {pages} {075101} (\bibinfo {year} {2016})}\BibitemShut
  {NoStop}%
\bibitem [{\citenamefont {Ro{\ss}nagel}\ \emph {et~al.}(2016)\citenamefont
  {Ro{\ss}nagel}, \citenamefont {Dawkins}, \citenamefont {Tolazzi},
  \citenamefont {Abah}, \citenamefont {Lutz}, \citenamefont {Schmidt-Kaler},\
  and\ \citenamefont {Singer}}]{Rosnagel2015}%
  \BibitemOpen
  \bibfield  {author} {\bibinfo {author} {\bibfnamefont {J.}~\bibnamefont
  {Ro{\ss}nagel}}, \bibinfo {author} {\bibfnamefont {S.~T.}\ \bibnamefont
  {Dawkins}}, \bibinfo {author} {\bibfnamefont {K.~N.}\ \bibnamefont
  {Tolazzi}}, \bibinfo {author} {\bibfnamefont {O.}~\bibnamefont {Abah}},
  \bibinfo {author} {\bibfnamefont {E.}~\bibnamefont {Lutz}}, \bibinfo {author}
  {\bibfnamefont {F.}~\bibnamefont {Schmidt-Kaler}}, \ and\ \bibinfo {author}
  {\bibfnamefont {K.}~\bibnamefont {Singer}},\ }\href
  {http://science.sciencemag.org/content/352/6283/325.abstract} {\bibfield
  {journal} {\bibinfo  {journal} {Science}\ }\textbf {\bibinfo {volume}
  {352}},\ \bibinfo {pages} {325} (\bibinfo {year} {2016})}\BibitemShut
  {NoStop}%
\bibitem [{\citenamefont {Abah}\ \emph {et~al.}(2012)\citenamefont {Abah},
  \citenamefont {Ro\ss{}nagel}, \citenamefont {Jacob}, \citenamefont {Deffner},
  \citenamefont {Schmidt-Kaler}, \citenamefont {Singer},\ and\ \citenamefont
  {Lutz}}]{PhysRevLett.109.203006}%
  \BibitemOpen
  \bibfield  {author} {\bibinfo {author} {\bibfnamefont {O.}~\bibnamefont
  {Abah}}, \bibinfo {author} {\bibfnamefont {J.}~\bibnamefont {Ro\ss{}nagel}},
  \bibinfo {author} {\bibfnamefont {G.}~\bibnamefont {Jacob}}, \bibinfo
  {author} {\bibfnamefont {S.}~\bibnamefont {Deffner}}, \bibinfo {author}
  {\bibfnamefont {F.}~\bibnamefont {Schmidt-Kaler}}, \bibinfo {author}
  {\bibfnamefont {K.}~\bibnamefont {Singer}}, \ and\ \bibinfo {author}
  {\bibfnamefont {E.}~\bibnamefont {Lutz}},\ }\href {\doibase
  10.1103/PhysRevLett.109.203006} {\bibfield  {journal} {\bibinfo  {journal}
  {Phys. Rev. Lett.}\ }\textbf {\bibinfo {volume} {109}},\ \bibinfo {pages}
  {203006} (\bibinfo {year} {2012})}\BibitemShut {NoStop}%
\bibitem [{\citenamefont {Jordan}\ \emph {et~al.}(2013)\citenamefont {Jordan},
  \citenamefont {Sothmann}, \citenamefont {S\'anchez},\ and\ \citenamefont
  {B\"uttiker}}]{PhysRevB.87.075312}%
  \BibitemOpen
  \bibfield  {author} {\bibinfo {author} {\bibfnamefont {A.~N.}\ \bibnamefont
  {Jordan}}, \bibinfo {author} {\bibfnamefont {B.}~\bibnamefont {Sothmann}},
  \bibinfo {author} {\bibfnamefont {R.}~\bibnamefont {S\'anchez}}, \ and\
  \bibinfo {author} {\bibfnamefont {M.}~\bibnamefont {B\"uttiker}},\ }\href
  {\doibase 10.1103/PhysRevB.87.075312} {\bibfield  {journal} {\bibinfo
  {journal} {Phys. Rev. B}\ }\textbf {\bibinfo {volume} {87}},\ \bibinfo
  {pages} {075312} (\bibinfo {year} {2013})}\BibitemShut {NoStop}%
\bibitem [{\citenamefont {Sothmann}\ \emph {et~al.}(2015)\citenamefont
  {Sothmann}, \citenamefont {S{\'a}nchez},\ and\ \citenamefont
  {Jordan}}]{0957-4484-26-3-032001}%
  \BibitemOpen
  \bibfield  {author} {\bibinfo {author} {\bibfnamefont {B.}~\bibnamefont
  {Sothmann}}, \bibinfo {author} {\bibfnamefont {R.}~\bibnamefont
  {S{\'a}nchez}}, \ and\ \bibinfo {author} {\bibfnamefont {A.~N.}\ \bibnamefont
  {Jordan}},\ }\href {http://stacks.iop.org/0957-4484/26/i=3/a=032001}
  {\bibfield  {journal} {\bibinfo  {journal} {Nanotechnology}\ }\textbf
  {\bibinfo {volume} {26}},\ \bibinfo {pages} {032001} (\bibinfo {year}
  {2015})}\BibitemShut {NoStop}%
\bibitem [{\citenamefont {Hofer}\ and\ \citenamefont
  {Sothmann}(2015)}]{PhysRevB.91.195406}%
  \BibitemOpen
  \bibfield  {author} {\bibinfo {author} {\bibfnamefont {P.~P.}\ \bibnamefont
  {Hofer}}\ and\ \bibinfo {author} {\bibfnamefont {B.}~\bibnamefont
  {Sothmann}},\ }\href {\doibase 10.1103/PhysRevB.91.195406} {\bibfield
  {journal} {\bibinfo  {journal} {Phys. Rev. B}\ }\textbf {\bibinfo {volume}
  {91}},\ \bibinfo {pages} {195406} (\bibinfo {year} {2015})}\BibitemShut
  {NoStop}%
\bibitem [{\citenamefont {Hofer}\ \emph {et~al.}(2016)\citenamefont {Hofer},
  \citenamefont {Souquet},\ and\ \citenamefont {Clerk}}]{PhysRevB.93.041418}%
  \BibitemOpen
  \bibfield  {author} {\bibinfo {author} {\bibfnamefont {P.~P.}\ \bibnamefont
  {Hofer}}, \bibinfo {author} {\bibfnamefont {J.-R.}\ \bibnamefont {Souquet}},
  \ and\ \bibinfo {author} {\bibfnamefont {A.~A.}\ \bibnamefont {Clerk}},\
  }\href {\doibase 10.1103/PhysRevB.93.041418} {\bibfield  {journal} {\bibinfo
  {journal} {Phys. Rev. B}\ }\textbf {\bibinfo {volume} {93}},\ \bibinfo
  {pages} {041418} (\bibinfo {year} {2016})}\BibitemShut {NoStop}%
\end{thebibliography}
%\bibliographystyle{apsrev4-1}
%%\bibliographystyle{aipauth4-1}

\end{document}